\documentclass[10pt,conference]{IEEEtran}
\usepackage{mathptmx} % This is Times font

\usepackage{fancyhdr}
\usepackage[normalem]{ulem}
\usepackage[hyphens]{url}
\usepackage[sort,nocompress]{cite}
\usepackage[final]{microtype}
\usepackage{flushend}
\usepackage[bookmarks=false]{hyperref}
\usepackage{setspace}
\usepackage{booktabs}
\usepackage{mathptmx}   
\usepackage{amsmath}
\usepackage{times}
\usepackage{threeparttable}
\usepackage[ruled,linesnumbered]{algorithm2e}
\usepackage{subfigure}
\usepackage{graphicx}
\usepackage{fancyhdr}
\usepackage[noend]{algpseudocode}
\usepackage[normalem]{ulem}
\usepackage[shortlabels]{enumitem}
\usepackage{lineno}
\usepackage{cite}
\usepackage{soul}
\soulregister\cite7
\soulregister\ref7
\soulregister\footnote7
\renewcommand{\thefootnote}{\fnsymbol{footnote}}

\title{Co-Optimizing Performance and Memory Footprint Via Integrated CPU/GPU Memory Management, an Implementation on Autonomous Driving Platform}
\author{\IEEEauthorblockN{Soroush Bateni\normalfont\textsuperscript{*}, Zhendong Wang\normalfont\textsuperscript{*}, Yuankun Zhu, Yang Hu, and Cong Liu}
\IEEEauthorblockA{\textit{The University of Texas at Dallas}
}
}

\begin{document}
\maketitle
\pagestyle{plain}
%\bibliographystyle{unsrt}
%\renewcommand{\baselinestretch}{1.18}
%\setlength{\baselineskip}{12pt}
%\setlength{\lineskiplimit}{1pt}
%\setlength{\parskip}{0pt}

%%%%%% -- PAPER CONTENT STARTS-- %%%%%%%%
\begin{abstract}
Cutting-edge embedded system applications, such as self-driving cars and unmanned drone software, are reliant on integrated CPU/GPU platforms for their DNNs-driven workload, such as perception and other highly parallel components. In this work, we set out to explore the hidden performance implication of GPU memory management methods of integrated CPU/GPU architecture. Through a series of experiments on micro-benchmarks and real-world workloads, we find that the performance under different memory management methods may vary according to application characteristics. Based on this observation, we develop a performance model that can predict system overhead for each memory management method based on application characteristics. Guided by the performance model, we further propose a runtime scheduler. By conducting per-task memory management policy switching and kernel overlapping, the scheduler can significantly relieve the system memory pressure and reduce the multitasking co-run response time. We have implemented and extensively evaluated our system prototype on the NVIDIA Jetson TX2, Drive PX2, and Xavier AGX platforms, using both Rodinia benchmark suite and two real-world case studies of drone software and autonomous driving software.

\end{abstract}
%\vspace{-2mm}
\section{Introduction}
\footnotetext[1]{These two authors contributed equally.}
%}

\renewcommand{\thefootnote}{\arabic{footnote}}

In recent years, machine learning (ML) applications, especially deep neural networks (DNNs), have penetrated into a wide range of edge devices and scenarios. Typical scenarios include the real-time object and pedestrian recognition of video streams from multiple cameras and object perceptions from high-volume LIDAR streams in autonomous driving and unmanned drones. Modern autonomous and edge intelligence systems are mostly deployed on resource-constrained embedded platforms that require performance and are restricted by the Size, Weight, and Power-consumption (SWaP) and cost. The hardware substrates of modern intelligent edge devices must seek the balance between accuracy, latency, and power budget. Currently, GPU is the most promising and widely used accelerator for the "autonomous everything" thanks to its throughput processing nature that matches the ML algorithms, which back the future autonomous applications. Specifically, the integrated CPU/GPU architecture is gaining increasing preference in embedded autonomous systems due to its programmability and the advantages in SWaP comparing to its discrete counterparts~\cite{jetson,smartcity}. To seize the trend, NVIDIA proposes its Jetson line of embedded platforms that target autonomous systems, and markets the Jetson as "the embedded platform for autonomous everything"~\cite{nvidiaembed}. Recently, Intel also launches the OpenVINO toolkit for the edge-based deep learning inference on its integrated HD GPUs~\cite{OPENVINO}.

Despite the advantages in SWaP features presented by the integrated CPU/GPU architecture, our community still lacks an in-depth understanding of the architectural and system behaviors of integrated GPU when emerging autonomous and edge intelligence workloads are executed, particularly in multi-tasking fashion. Specifically, in this paper we set out to explore the performance implications exposed by various GPU memory management (MM) methods of the integrated CPU/GPU architecture. The reason we focus on the performance impacts of GPU MM methods are two-fold. First, emerging GPU programming frameworks such as CUDA and OpenCL support various MM methods for the integrated CPU/GPU system to simplify programmability. However, these methods are encapsulated in the runtime library/GPU drivers and are transparent to programmers. It is still unclear how to adopt the GPU MM method that best-matches single task and co-run tasks in autonomous workloads, such that memory footprint and latency performance could be optimized. Second, the integrated GPU system distinguishes itself from discrete GPU mainly by employing a shared physical memory pool for CPU and GPU. This may result in challenges of performance interference, GPU resource utilization, and memory footprint management considering key components of autonomous workloads are memory-intensive. 

To fully disclose the performance implications of GPU MM methods to the emerging autonomous workloads, we conduct a comprehensive characterization of three typical GPU MM methods (i.e., Device Memory, Managed Memory, Host-Pinned Memory) that are applicable to all integrated CPU/GPU architectures (e.g., those provided by NVIDIA CUDA and AMD OpenCL frameworks), through running both microbenchmarks (Rodinia) and real autonomous system workloads (drone system~\cite{djidrone2017sensing} and Autoware~\cite{autoware} for autonomous driving) on top of NVIDIA Jetson TX2/PX2 and Jetson AGX Xavier platforms, respectively. Our characterization reveals several key observations and corresponding design opportunities. 

First, we observe that different GPU MM methods can lead to different GPU response times and GPU utilization for given application. It is non-trivial to determine the GPU MM method that achieves the best performance considering the autonomous driving workloads are becoming even more complex. This motivates us to propose a light-weight analytical performance model that quantifies the performance overheads of different GPU MM methods and further eases the GPU MM method selection. Second, emerging autonomous driving and drone software consist of various key components such as object detection (perception module) and LIDAR-based localization. These DNNs-based recognition functions could be extremely memory-consuming. The concurrent multi-tasking, which is a typical scenario in embedded autonomous environments, will further worsen this situation. This motivates us to also consider the memory footprint as a critical factor when choosing the GPU MM methods. 

Based on these two basic observations, we set out to explore the co-execution cases where multiple concurrent tasks are executed with different GPU MM methods. We observe that by strategically assigning the co-executed tasks with specific GPU MM methods, the multi-tasking environment provides a counter-intuitive opportunity to significantly minimize the memory footprint of the overall system without sacrificing the task processing latency, sometimes even with reducing the overall latency. This implies that two goals, the reductions of both system-level memory footprint and per-task latency, could be cooperatively achieved by simply grouping tasks and assigning per-task memory management method. Note that we use memory footprint/usage interchangeably in this paper. This motivates us to 
propose a runtime scheduler that exploits the aforementioned co-execution benefit on GPU by ordering the incoming tasks and assigning GPU MM methods. 

We extensively evaluate our prototype on the NVIDIA Jetson TX2, Drive PX2, and Xavier AGX platforms, using both the Rodinia benchmark suite and two real-world case studies of drone software and autonomous driving software. Results demonstrate that the analytical performance model is sufficiently accurate, which gains the average error rate of 9.3\% for TX2 and 9.3\% for AGX. When applying to real-world applications, our solution can significantly reduce memory usage of iGPU on average by 50\% for Autoware on PX2 and by 69.0\%/64.5\% for drone obstacle detection ON TX2/AGX; and improve response time performance on average by 10\% for Autoware on PX2 and by 58.9\%/11.2\% for drone obstacle detection on TX2/AGX.
This paper makes the following contributions:
\begin{itemize}
\vspace{-1mm}
    \item We conduct a comprehensive characterization of three typical GPU memory management  methods on a variety of workloads including microbenchmarks and real autonomous workloads.%\vspace{-2mm}
    \item We propose an analytical performance model that is able to estimate the system response time for a certain task or kernel under any GPU memory management method. %\vspace{-2mm}
    \item We propose a memory footprint minimization mechanism and a runtime scheduler. Together, they make smart decisions to minimize the memory footprint of applications and make certain that response performance will not be noticeably affected and may be enhanced in many cases.%\vspace{-2mm}
    \item We implement our design as a system prototype and extensively evaluate our design using two real autonomous systems: a drone system and the Autoware autonomous driving system.%\vspace{-1mm}
\end{itemize}
%\vspace{-2mm}
\section{Background}
%\vspace{-1mm}

%\vspace{-1mm}
\subsection{Integrated GPU design}
\vspace{-1mm}
As we discussed earlier, the recent trend in compute and data-intensive embedded systems has resulted in the rise of powerful integrated GPU (iGPU) in embedded chip design. 
iGPU in embedded systems provides a reasonable trade-off between  performance, energy usage, thermal capacity, and space usage compared to a discrete GPU component~\cite{jetson}. To this end, various manufacturers have begun to offer GPU-augmented multicore platforms specifically catered to embedded use cases such as autonomous drones and vehicles. Previous studies have shown that iGPU might be sufficient for workloads of autonomous drones and vehicles if resources are well managed~\cite{smartcity,sor1,sor2}. 

In a traditional multi-cores architecture with discrete GPUs, a discrete GPU is handled semi-autonomously with a separate physical memory specifically allocated to it. However, rather than having a separate, high-speed/high-bandwidth memory, iGPU is usually connected to the chip interconnect and thus compete with CPU for memory resources. 

Among architectures featuring iGPU, such as Intel SVM~\cite{george2011technology} and AMD HUMA~\cite{brookwood2010amd}, the most prominent  ones are designed by NVIDIA, including the Parker and Xavier System-on-Chips (SoC), which are collectively called the Tegra family of SoCs. These SoCs include a powerful Pascal or Volta-based GPU, and are practically used in many autonomous drone or vehicle applications~\cite{jetson}. Next, we give an overview of NVIDIA Parker and Xavier, which are the focused integrated CPU/GPU architectures of this paper.

\begin{figure}[t]
\centering
           \includegraphics[width=0.9\linewidth]{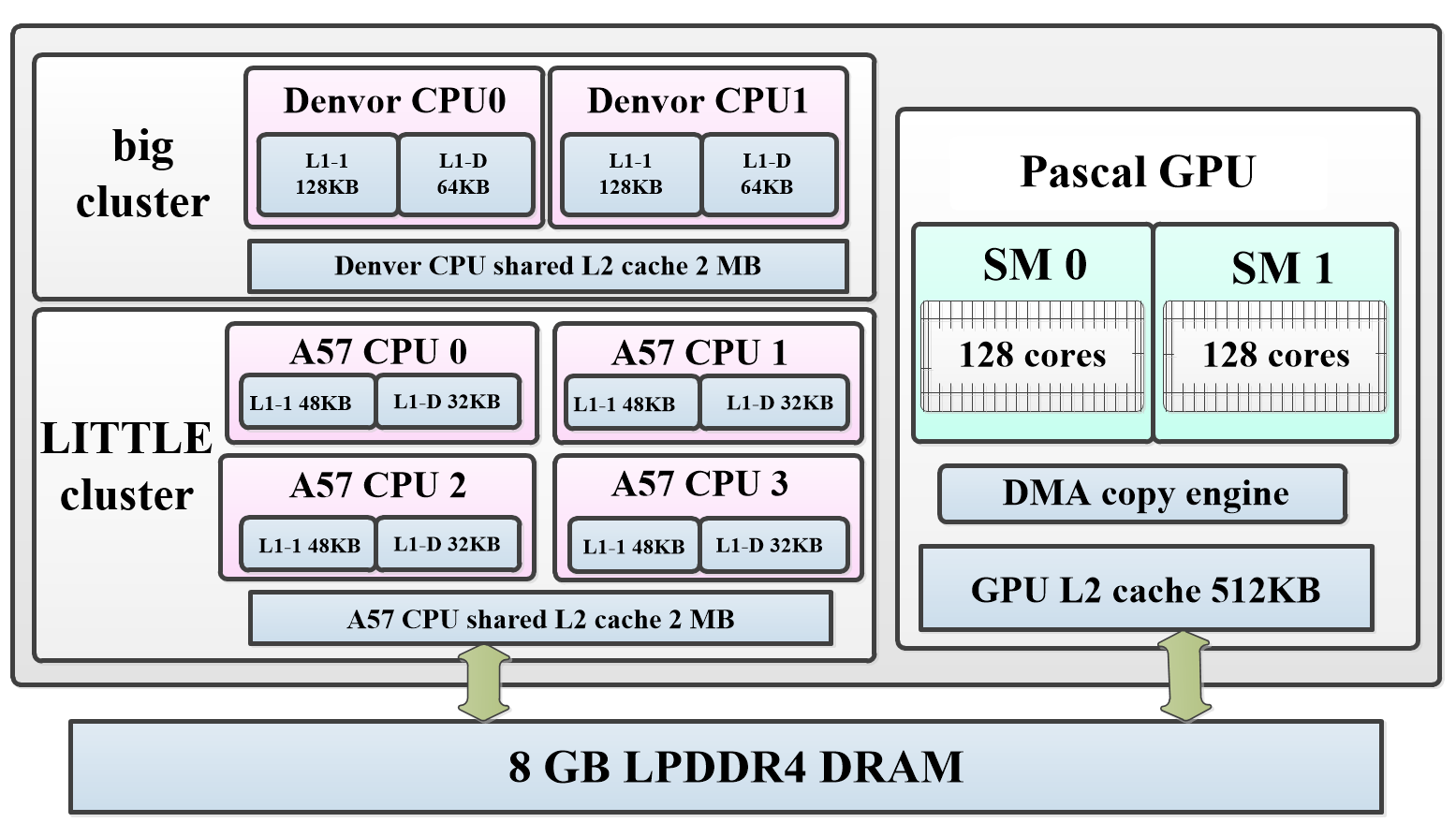}
            \vspace{-4mm}
            \caption{Overview of Parker-based TX2 module.}\label{tx2}
            \vspace{-4mm}
\end{figure}

\noindent\textbf{NVIDIA Parker SoC.} 
Fig.~\ref{tx2} illustrates the NVIDIA Parker SoC used in platforms such as Jetson TX2 and NVIDIA Drive PX2\footnote{NVIDIA PX2 uses two Parker-based chips (i.e., TX2) that are interconnected over Ethernet, plus two discrete Pascal GPUs on board.}, which are currently adopted by many automotive manufacturers (e.g., used in Volvo XC90 and Tesla Model S). 
NVIDIA Parker SoC consists of two Denver2 and four ARM Cortex A57 cores. The chip has an integrated two-core Pascal GPU with 256 CUDA cores connected via an internal bus. The iGPU and CPU share up to 8 GB memory with up to 50 GB/s of bandwidth in typical applications of NVIDIA Parker. 

 \normalsize

\vspace{1mm}\noindent\textbf{NVIDIA Xavier SoC.} NVIDIA Xavier is the latest SoC in the NVIDIA AGX Xavier and Drive AGX Pegasus platforms. Xavier has an 8-core "Caramel" CPU (based on ARM V8) and an 8-core Volta GPU with 512 CUDA cores, and 16GBs of shared main memory 
\footnote{Note that even though our motivation and evaluation are based on NVIDIA Parker and Xavier, our methodology is applicable to all integrated CPU/GPU architectures that utilize a unified memory architecture.}.

\noindent\textbf{GPU terminology.} 
We define keywords used throughout this paper that pertain to the integrated CPU/GPU architecture and GPUs in general. \textbf{Host} is the designated term given to the CPU and the supporting architecture that enables memory and instructions to be sent to GPU. \textbf{Device} is the equivalent of GPU on iGPU systems. 
\textbf{Kernel} is the collective name for the logical group of instructions sent to GPU for execution. Kernels are set-up by the developer and consist of threads, thread-blocks, and grids. \textbf{Memory} is a crucial part of kernel execution. Prior to execution, data has to be introduced to the GPU in some way. After the kernel finishes the execution, the data needs to be sent back, or the host needs to be informed depending on the memory management  technique.\\
\vspace{-4mm}
\subsection{Memory Management (MM) Methods} \label{sec:MM_motiv}

There are various methods to manage a unified memory between CPU and iGPU. 
All existing memory management techniques can be classified into three categories~\cite{dashti2017analyzing}: Device Memory (D), Managed Memory (M), and Host-Pinned Memory (H). There are some internal differences among these techniques based on their handling of cache coherency and data communication between host and device.

\noindent \textbf{Device Memory (D).}
Under D, data should be copied to the allocated memory region for the device from host prior to GPU computation and has to be copied back to the host memory region after GPU computation completes. This procedure is called data transfer, which mainly contributes to the overhead when the host communicates with the device. Besides, before data transferring, a comparable data-size space has to be created first on both host and device memory regions, which requires 2x memory space to be allocated to the same data. For example, the $CudaMalloc()$ API corresponds to this behavior and is the default method used in CUDA to allocate memory on the device. On an embedded platform, this extra memory usage cannot be ignored since the integrated memory size is highly constrained.

\noindent \textbf{Managed Memory (M).} 
When policy M is applied to allocate memory in an integrated architecture, the system will create a pool of managed memory, which is shared between the CPU and GPU. 
For example, $CudaMallocManaged()$ corresponds to this behavior, an API that was first introduced in CUDA 6.0 to simplify memory management. The pool is allocated on the host and it will be mapped to the device memory address space such that the GPU can seamlessly access the region rather than explicitly transfer data. 
Although policy M can allocate a unified memory on a platform such as Parker to eliminate the explicit data transfer, it will nonetheless incur some extra overhead due to maintaining data coherency. Since the data in the pool can be cached on both CPU and GPU, cache flushing is required in certain scenarios.
More importantly, GPU and CPU are not allowed to access the data allocated in the managed memory concurrently to guarantee coherency. 

\noindent\textbf{Host-Pinned Memory (H).} Policy H can avoid memory copy between CPU and GPU similar to M. The main difference between them lies in handling the cache coherency issues. In H, both GPU and CPU caches will be bypassed to ensure a consistent view. 
As a result, the CPU and GPU are allowed to access the same memory space concurrently, which is forbidden under M. This policy can be invoked by using the $CudaHostAlloc()$ API in 
CUDA programming.

%\vspace{-1mm}
\section{Motivation}
%\vspace{-1mm}

In this section, we present three fundamental observations that have motivated our design of efficient memory management on the integrated CPU/GPU architecture for modern autonomous driving applications.

\subsection{Performance under GPU MM Policies} \label{sec:perf_MM}
%\vspace{-1mm}
To provide better memory allocation approaches for the modern autonomous driving-targeted integrated CPU/GPU architecture, a critical first step is to understand the performance (e.g. GPU time) and resource consumption (e.g. memory footprint and GPU utilization) under different GPU memory management policies.

\begin{figure}[t]
\centering
            \includegraphics[width=8.5cm,height=2.8cm]{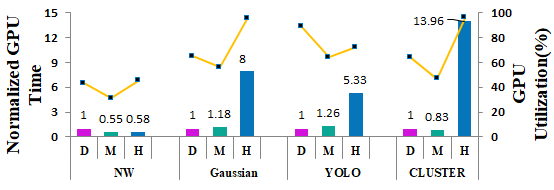}
            \vspace{-4mm}
            \caption{GPU time and utilization of tested applications.}
           \label{GPUutilizationinmotivation}
\vspace{-6mm}
\end{figure}

To that end, we run a set of experiments on top of the Parker-based Jetson TX2 platform~\cite{franklin2017nvidia} 
using assorted Rodinia benchmarks~\cite{rodinia}, YOLO-based object detection for camera perception~\cite{joseph2016yolo}, and Euclidean clustering for LIDAR-based object detection~\cite{7368032}. Fig.~\ref{GPUutilizationinmotivation} reports the measured GPU time of the tested benchmarks, including \textit{NW}, \textit{Gaussian}, YOLO, and clustering instance on TX2 under various MM policies. 
We use \textit{gettimeofday()} to extract the time of memory copies and kernel executions. The overhead of function calls is not considered and there is no memory copies considered in the GPU time for M/H policy.
Typically, the run time of a task consists of CPU time on pre-processing and GPU time. Considering the CPU time is quite consistent and negligible compared to GPU time in a computation-intensive task, we focus on the GPU time in this characterization. Specifically, the GPU time includes transfer delay and kernel computation time under D policy, while the GPU time only includes the kernel computation time under M/H policy.

We can observe that there is no clear winner between these GPU MM policies in terms of GPU time and utilization. In most cases, the H policy leads to the worst performance, e.g., for Gaussian, YOLO and CLUSTER, due to the fact that H disables both CPU-/GPU-side cache and incurs full memory access latency %\textcolor{red}{\sout{for each memory access}}. 
Considering D and M policy, D yields the best performance for Gaussian while yielding the worst performance for NW. Since D incurs extra data transfer cost between host-allocated memory and device-allocated memory, the cost will be exacerbated when application data size is large (e.g., Gaussian with small data vs. NW with large data). We can also observe the same trend in YOLO and CLUSTER instance, as CLUSTER has larger data size. Meanwhile, M needs to flush the cache frequently to maintain coherency at critical time instances such as kernel launch. Both Gaussian and YOLO are particularly hit by this overhead since they involve many kernel launches.

We also measure the GPU utilization using the modified tool, tx-utils \cite{tx-utils}, when these benchmarks execute under different MM policies, which is depicted as lines in Figure \ref{GPUutilizationinmotivation}. 
We observe that, the GPU is significantly underutilized under M method. For example, for NW and CLUSTER, the utilization is even less than 50\%. In comparison, under D and H policy, the GPU utilization is relatively high. By further analysis, we find that, under M policy, the GPU is still inactive though a kernel has launched on GPU side.

\noindent\textbf{Observation 1:} %We observe the trend that 
Each memory management policy may yield a completely different performance for different applications. Particularly, the GPU is significantly underutilized under M policy. There still lacks a clear understanding surrounding the trade-off behind each policy and how such trade-off would apply to different applications.

\begin{table}[t]
	\begin{minipage}[b]{0.545\linewidth}
		\includegraphics[width=\linewidth]{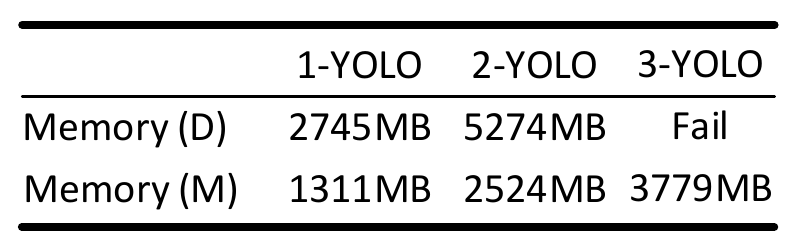}
		\vspace{-4mm}
		\caption{Peak memory usage of running YOLOs on TX2.}
		\label{fig:peakmemory} %-----------------
%		\vspace{-0.5cm}
	\end{minipage}%
	\begin{minipage}[b]{0.43\linewidth}
		\includegraphics[width=\linewidth]{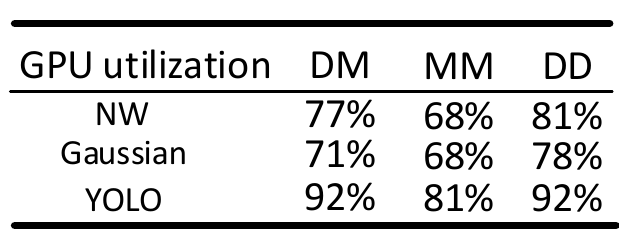}
		\vspace{-4mm}
		\caption{GPU utilization of test applications.} \label{fig:gpuutilization}
	%	\vspace{-0.5cm}%------------- 
	\end{minipage}%
	\vspace{-8mm}
\end{table}

\begin{figure}[t]

\subfigure[Memory usage of NW, Gaussian and YOLO.]{

\centering
           \includegraphics[width=0.45\linewidth]{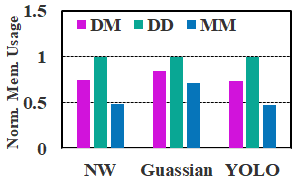}
            \label{rodiniamem2}
}
\subfigure[Normalized GPU time of NW, Gaussian and YOLO.]{

            \includegraphics[width=0.45\linewidth]{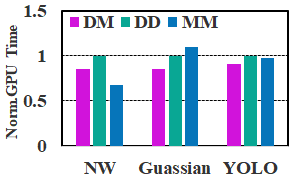}
            \label{rodiniatime2}
}
\vspace{-2mm}

\caption{Normalized memory usage and GPU time of two instances of benchmarks and YOLOs.}
\label{Fig.Rodinia}
\vspace{-6mm}
\end{figure}
\vspace{-1mm}
\subsection{Memory Pressure} \label{sec:mem_pres}
%\vspace{-1mm}
A characteristic of systems with iGPU
is a limited memory pool shared between CPU and GPU. 
In such applications as DNNs ~\cite{zhou2018s}, this limited pool can become a real resource bottleneck. 
Table~\ref{fig:peakmemory} showcases the peak memory usage of one to three YOLO instances~\cite{joseph2016yolo} under D and M policy. 
We observe that running YOLO may consume more than 38.6\% of the total shared memory for a single instance and can fail when 3 instances simultaneously execute under D policy. However, it is clear that switching from D to the M policy can result in significant memory footprint reduction and accommodate one more instance. 

\noindent\textbf{Observation 2:} Considering performance alone might not be sufficient for achieving high throughput and low-latency on integrated CPU/GPU architecture. Any proposed method which determines the memory management policy should contain precautions to minimize memory footprint.

\subsection{Opportunities under Multitasking} \label{sec:k_co}
%\vspace{-1mm}
Considering modern autonomous vehicle or drone can carry multiple cameras, where each camera relies on individual YOLO instance, it is common that multi-tasks co-execute simultaneously in the system. 
Therefore, we co-execute micro-benchmarks and YOLO instances to explore the chances to further optimize memory usage under multi-tasks scenarios. 

Interestingly, we observe a counter-intuitive opportunity under multitasking scenarios, where it may be possible to minimize memory footprint without sacrificing (even with improving) performance. Fig.~\ref{Fig.Rodinia} depicts the result of our motivational experiments,
including the memory footprint and GPU time of two co-executed NWs, Gaussians and YOLOs.  

Fig.~\ref{rodiniamem2} depicts the normalized memory footprint of two instances of NW, Gaussian and YOLO under various MM policies (e.g., D/M indicates that one benchmark is executed under D policy and the other one under M policy). Evidently, M/M occupies less memory than D/D and D/M on all benchmarks. This is due to the fact that M eliminates the need for additional memory copy. Thus, maximizing the number of applications that use the M policy can result in the best memory savings.

However, the latency, indicated by GPU time, depicts a different story, as is shown in Fig.~\ref{rodiniatime2}. Only for NW benchmark can the mixed M/M policy yield the best performance. In comparison, the mixed D/M policy actually achieves the best possible latency performance for both Gaussian and YOLO. More importantly, we consider the co-execution benefit in terms of GPU utilization as shown in Table.~\ref{fig:gpuutilization}. Obviously, the M/M policy always leads to the under-utilization of GPU compared to either mixed D/M or D/D policy. Under D/M policy, both Gaussian and YOLO not only reduce the memory usage but also benefit the performance compared to D/D policy, hitting two optimization goals simultaneously.

\begin{figure}[t]
\centering
\includegraphics[width=8.5cm, height=3.5cm]{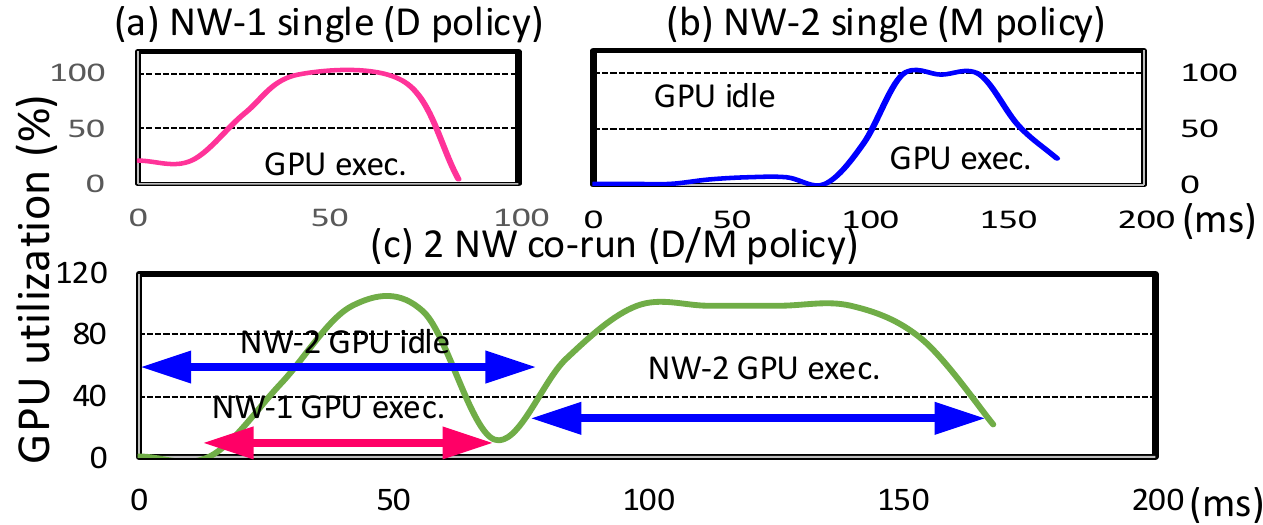}
\vspace{-4mm}
\caption{Illustration of the overlapping potential.}
\vspace{-6mm}
\label{fig:interleave}
\end{figure}

\noindent\textbf{Opportunities:} Although the performance results in Fig.~\ref{Fig.Rodinia} may at the first glance appear disorderly, they follow a predictable pattern. We report the detailed GPU utilization along the run time of NW kernel under D, M and D/M co-run, as shown in Fig.~\ref{fig:interleave}(a), (b), and (c). For the M policy, the first data access executed in GPU will initiate a memory mapping procedure that stalls the GPU (shown as an idle period). Based on our observation, the mapping typically occurs at the beginning of the kernel execution and would last for a while; that's why the GPU is still inactive though the kernel has launches on GPU. Furthermore, M takes high cost to maintain data coherency, while H directly disables cache on CPU and GPU sides. Both cases may incur large latency overhead in practice, particularly for those cache-intensive workloads.

Fig.~\ref{fig:interleave}(c) depicts the co-execution schedule under the D/M policy. When kernel $NW_2$ stalls for page mapping under the M policy, another kernel $NW_1$ using the D policy can switch in and execute during the idle of $NW_2$. Theoretically, D/H is similar to D/M because H and M have similar overhead structures for the idle period. However, based on our observations, H can significantly degrade an application's latency performance. D/H is thus not preferred in practice.

\begin{figure}[t]
\centering
\includegraphics[width=1.05\linewidth]{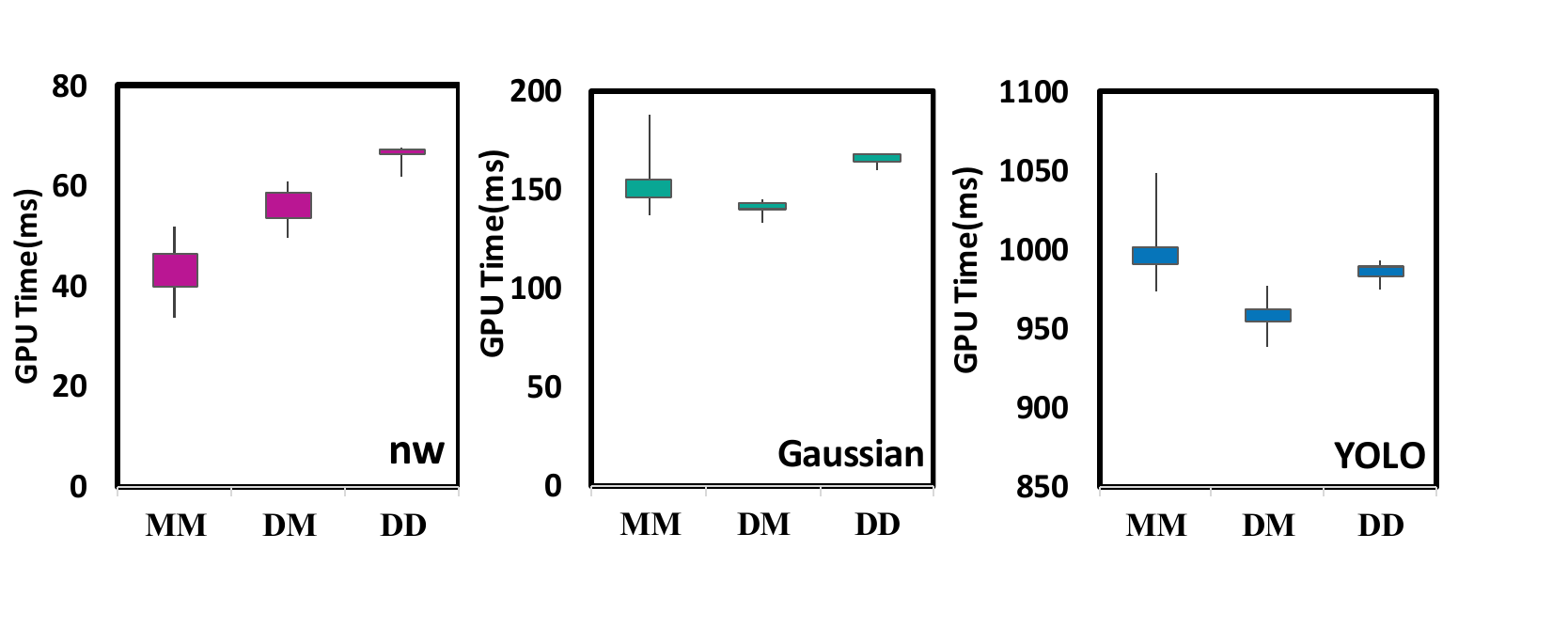}
\vspace{-13mm}
\caption{GPU time distribution of DD, DM and MM.}
\vspace{-5mm}
\label{fig:distribution}
\end{figure}

Note that such overlapping benefit may not apply to D/D or M/M co-run. Considering GPU typically serializes multi-kernel operations by default, the D/D co-execution could be ruled out due to its non-preemption property and data copy overheads. For M policy, the address mapping and cache flushing stage involve complex driver activities, which have not been disclosed by NVidia. \textbf{\textit{Simply co-running kernels 
%\sout{by default} 
using M/M policy not only leads to both kernels' performance unpredictable due to the interference of driver activities, but also results in significant GPU under-utilization}}. To depict the inefficiency and instability of M/M co-run, we report the average GPU utilization of NW, Gaussian, and YOLO under D/M, M/M, and D/D policies in Table~\ref{fig:gpuutilization}, and report the GPU time distribution of three tasks 
%\textcolor{red}{\sout{under three GPU MM policies}} 
by repeating 10 times in Fig.~\ref{fig:distribution}. We observe that the M/M policies co-run always comes with the lowest GPU utilization and largest spread.

\noindent\textbf{Observation 3:} In multitasking scenarios, there are opportunities for enhancing performance through assigning different memory management policies to different applications to enable execution overlapping. \textit{A strategic assignment may enable performance and memory usage reduction as well as GPU utilization improvement simultaneously}.

%\vspace{-2mm}
\section{Design}
\label{sec:design}
%\vspace{-1mm}
In this section, we propose a memory management framework for the integrated CPU/GPU architecture, which identifies the best GPU MM policy for each application with the goal of optimizing overall response time performance with minimum memory footprint for applications running on the iGPU. Our design consists of two key components. First, we establish an analytical performance model to estimate the response time of any given application under the three GPU MM policies. 
Second, the estimated results are used as the initial guidance in a task scheduling process. This process decides the final GPU MM policy assignment for each application as well as whether two kernels shall be co-executed to benefit from overlapping. 

\subsection{Analytical Model} \label{sec:ana_model}
%\vspace{-1mm}
To wisely decide on the memory management policy for an application that results in the best response time performance, we establish an analytical performance model to estimate the potential overhead caused by each policy for that application. For each MM policy, the associated overhead for running a single application on GPU is defined to be the total GPU response time (including all memory management procedures) of the application minus the pure GPU execution time of that application. 
Table~\ref{tbl:params} lists our identified parameters that are crucial to construct the performance model for each MM policy, along with their description. 
Next, we detail our performance modeling for each memory management policy.

\begin{table}[]
\centering
%\footnotesize{
%\setlength{\tabcolsep}{7mm}
\resizebox{\columnwidth}{30mm}{
\begin{tabular}{@{}llll@{}}
\toprule
Symbol        & Unit      & Description                       	&Source\\ \midrule
$B_{dh}$      & byte      & Device to host data size        	& Profiling\\
$B_{hd}$      & byte      & Host to device data size        	& Profiling	\\
$S_{i}$       & byte      & Single kernel size			   		& Profiling\\
$N_{k}$       & -         & Kernel count                   		& Profiling\\
$N_{c}$       & -         & Kernels launched  w/ dirty cache         &Profiling\\
$N_{m}$       & -         & Memory copy  calls count       		& NVIDIA Nsight \cite{nsight20133}\\
$N_{L2}$      & -         & GPU L2 cache Instructions          	& NVIDIA Nsight \\
$\tau$        & \% & GPU L2 cache hit rate           	& NVIDIA Nsight\\
$l_{hd}$      & ms/byte   & Host to device latency         		& Benchmarking\\
$l_{dh}$      & ms/byte   & Device to host latency          	& Benchmarking\\
$l_{mapping}$ & ms/byte   & GPU Page mapping latency			& Benchmarking\\
$l_{ini}$       & ms        & Host memory init. latency 		& Benchmarking\\
$l_{L2}$      & ms        & GPU L2 cache access latency     	& Benchmarking\\
$l_{mem}$     & ms        & GPU memory access latency      		& Benchmarking\\ 
$Tr_{ini}$       & ms  	  & Data transfer startup time      	& Benchmarking\\
$Ca_{cpu}$       & ms  	  & CPU cache flush time   		    	& Benchmarking\\
$Ca_{gpu}$       & ms  	  & GPU cache flush time   		    	& Benchmarking\\ \bottomrule
\end{tabular}
%the second bracket for resizebox
}
%}
\vspace{0mm}
\caption{Measurable parameters for building the model.}
\label{tbl:params}
\vspace{-8mm}
\end{table}\normalsize

\noindent \textbf{Device Memory (D).} For D policy, the overhead mainly comes from the communication between CPU and GPU, during which data transfers from host to device and then is copied back from device to host. Although the CPU and GPU share the same memory, the memory transfer is still inevitable under D allocation policy. Differing from discrete GPU systems, which need to transfer data from the CPU main memory to the device memory, iGPU systems just transfer data from one region of main memory to another region allocated for GPU usage. Considering that the entire process still needs to use the Direct Memory Access (DMA) copy engine, which is similar to the process on discrete systems, the prediction model of memory copy time of discrete GPU can also work for iGPU. Therefore, it is reasonable to apply the LOGGP performance model as the basis of our transfer time model \cite{alexandrov1995loggp,van2014performance}.

LOGGP is a state-of-the-art performance model that models point-to-point communication time of short fixed-size messages using the following parameters: 
latency ($L$) is the communication latency from the source processor to the target processor, process time ($o$) is the process time taken by the processors, 
bandwidth ($G$) represents the bandwidth for long messages. In a scenario where processor A wants to send a $k$ byte message to processor B, In the beginning, processor A has to take $o$ time to start the transfer and $G*k$ time to send out the message, then, $L$ time will be spent on communicating with B, and finally, processor B will take $o$ time to process the received message. 
To sum them up, the total communication time can be represented as~\cite{alexandrov1995loggp}:
%\vspace{-2mm}
\begin{equation} \label{equ:LOGGP}
%\vspace{-2mm}
\text{Communication time} = L+o+G*k+o.
\end{equation}

On integrated CPU/GPU architecture, data transfer can be treated as sending fixed-size messages. Thus, Equ.~\ref{equ:LOGGP} can serve as a foundation to calculate the transfer time under D, which is the sole overhead incurred under this policy. 
Differently, the data is directly sent from one memory region to another. Thus, the second $o$ in Equ.~\ref{equ:LOGGP} can be eliminated for calculating the data transfer time in this case ~\cite{van2014performance}, considering there is no process time involved upon receiving the data. Moreover, the transfer speed $G$ in Equ.~\ref{equ:LOGGP} is dependent on the direction of data travel (in communication the speed is usually assumed to be symmetric). In other words, the value of $G$ differs depending on whether the data is copied from host to device or it is copied from device to host. Finally, it is important to note that memory initialization overhead can also be a factor when data is copied back from device to host. This overhead interestingly is only present if the destination in the host region is being accessed for the first time. This overhead results from the lazy memory allocation nature of the operating system (Linux in our case)~\cite{love2005linux}. 

Using our identified variables in Table~\ref{tbl:params}, we define the transfer time denoted $O_D$ (thus the overhead) under the D policy as follows\footnote{Here we only consider the synchronized transfer overhead, for example for when the default CUDA memory copy function $cudamemcpy$ is invoked.}:
%\vspace{-2mm}
\begin{equation}\label{equ:M1}
{O_{D}=N_{m}*Tr_{ini}+(l_{dh}+l_{ini})*B_{dh}+l_{hd}*B_{hd}}
%\vspace{-2mm}
\end{equation}

\noindent{in which $N_{m}$ denotes the number of memory copy occurrences, and $Tr_{ini}$ is a replacement of $L+o$ in Equ.\ref{equ:LOGGP} to represent the startup time, which is a hardware-dependent constant. The rest of Equ.\ref{equ:M1} is used to represent the time it takes to transfer data (which was denoted by G*k in Equ.\ref{equ:LOGGP}). $l_{ini}$ is the variable for memory initialization in Equ.~\ref{equ:M1}; $l_{ini}$ is zero if the data is already allocated.}

\noindent\textbf{Managed Allocation (M).} Under M, there is no data transfer between host and device for iGPU. 
In iGPU systems, the memory copy is instead replaced by address mapping and cache maintenance operations which happen at the start and completion time of a kernel~\cite{nvidia2014toolkit} to guarantee a consistent view for host and device.
Next, we will give a detailed explanation for the overhead of the mapping and maintenance operations, which are the sole source of the overhead under M policy.

\noindent{\textit{Kernel Start}.}
Whenever an allocated managed region is modified by the host under M policy, the corresponding data in the unified main memory becomes outdated (without a cache flush). 
In this case, if a kernel (denoted by $K_i \in K$, $K$ being the set of a kernels in this scenario) starts executing on the GPU and initiates memory access to this allocated memory region under M, a cache flushing between host and device will need to be enforced to ensure data consistency (host cache flush). This cache operation has to be done before GPU can actually access the data. 

Thus, the intuition behind calculating the overhead under M is to deduce the latency of page allocation and mapping on GPU, and cache flushing on CPU. We use $T_{l}$ to represent the overhead during kernel launch:

%\vspace{-2mm}
\begin{equation}\label{Tl}
%\vspace{-2mm}
T_{l}=\sum\nolimits_{i=1}^{N_{c}} (S_{i}*l_{mapping}+Ca_{cpu}),
\end{equation}
in which $S_{i}$ represents the accessed data size by $K_{i}$ (that needs to be mapped from host to device memory), $N_{c}$ represents the size of set K, and $l_{mapping}$ represents the time cost per byte to create and map one page on GPU. $Ca_{cpu}$ represents the overhead of cache flushing from host to device. We regard this kind of overhead as a constant which is calculated as the time required to flush all the CPU cache into main memory (to provide an upper bound).

\noindent{\textit{Kernel completion}.}
During  kernel execution, the managed data is assumed to be local to GPU. This is done via disabling CPU access to this memory region while the kernel is executing. During kernel completion, if the data in the managed region was accessed by the kernel on GPU, it would be considered outdated to CPU because of memory caching. To ensure data consistency, the modified data present in GPU cache must be written back to the main memory after the kernel is completed. We use $T_{s}$ to represent the overhead caused by this required synchronization. We assume all cache lines for the accessed data will be out of date after a kernel is completed to account for the worst-case scenario, and the overhead of writing all the data back to memory could be represented as a constant $Ca_{gpu}$. Thus, $T_{s}$ can be derived as:
%\vspace{-1mm}
\begin{equation}
%\vspace{-1mm}
T_{s}=\sum\nolimits_{i=1}^{N_{k}}(B^i_{write}*l_{mem})\leq N_{k}*Ca_{gpu},
\end{equation}\label{T_{s}}
in which $B^i_{write}$ indicates the modified data size by Kernel i, and $N_{k}$ indicates the kernel count in an application.

The overhead of the application under the M policy ($O_{M}$) can be calculated as the summation of $T_{l}$ and $T_{s}$:
%\vspace{-1mm}
\begin{equation}\label{equ:M2}
%%\vspace{-1mm}
O_{M}=T_{l}+T_{s},
%\vspace{-1mm}
\end{equation}
%%\vspace{-1mm}
\noindent\textbf{Host-Pinned Allocation (H).}
For H, the overhead comes from the penalty of the disabled cache. Under the other two policies, global data could be cached in GPU and CPU's L2 cache. However, the memory region allocated by H is marked as "uncacheable"~\cite{nvidia2014toolkit}, thus, all access to this region will bypass cache and go to the main memory directly. Hence, the intuition behind calculating the overhead of H is to get the sum of latency penalties of all memory accesses compared to the cached scenario, which is given by: 
%\vspace{-2mm}
\begin{equation}\label{T_3}
%\vspace{-2mm}
   T_{c}= \sum\nolimits_{i=1}^{|K|}(\tau*N_{L2}*\overline{penalty})
\end{equation}
The penalty of cache miss denoted by $T_c$ can be given by:
%\vspace{-1mm}
\begin{equation}
%\vspace{-1mm}
\textit{$penalty_{i}$}\ =\left\{
\begin{array}{rcl}
0  && {IPL*TPL<max_{i}} \\
(l_{mem}-l_{Gcache}) & & {Otherwise}\\
\end{array} \right.
\end{equation}
\noindent where $l_{mem}$ and $l_{Gcache}$ represent the average instruction access latency of the main memory and the L2 cache respectively. The penalty of L2 cache miss is represented as the difference between $l_{mem}$ and $l_{Gcache}$.  However, the memory access latency could be hidden by the instruction pipeline~\cite{resios2011gpu}, implying that this latency penalty is zero. IPL and TPL represent the Instruction Level Parallelism and Thread Level Parallelism, and $max_{i}$ represents the Parallelism at peak throughput, which is measured from ~\cite{resios2011gpu}. $\tau$ represents the total hit rate of GPU L2 cache obtained under D, which will be the potential overhead under H (obtained from the performance counter), and $N_{L2}$ represents the number of instructions that access the GPU L2 cache. Moreover, H requires the same page allocation and mapping procedures as the M policy. We use ${O_{H}}$ to denote the total overhead under the H policy, and calculate it as:
%\vspace{-1mm}
\begin{equation}\label{M3}
%%\vspace{-2mm}
O_{H}=T_{c}+\sum\nolimits_{i=1}^{N_{k}} S_{i}*l_{mapping},
%\vspace{-1mm}
\end{equation}
\noindent\textbf{Summarized guidelines on MM selection.} From our detailed performance models, we observe that the overhead of D policy (Equ.~\ref{equ:M1}) heavily depends on $B_{dh}$ and $B_{hd}$, which means that applications with large data will be negatively affected by D. Moreover, the overhead of M (Equ.~\ref{equ:M2}) also depends on data size ($B_i$). Nonetheless, the data-dependent overhead (mainly from mapping) is much smaller compared to D because there is no need to physically move the data. However, M-induced overhead also depends on $N_k$, which indicates that applications with a lot of kernel launches (e.g., $Gaussian$) will be negatively affected. Finally, we note that the overhead of H (Equ.~\ref{M3}) depends on $\tau$ and $N_{L2}$, which indicates, if an application benefits heavily from cache, its performance will be negatively impacted under H policy. Besides, D or M is mostly preferred in practice though H may yield an overhead similar to D or M, because the disabled CPU cache under H will lead to further performance degradation for workloads executing on the CPU side.

Although our implementation (restricted by the hardware platform we can obtain and the software components we can access) only presents a single case of autonomous platforms, our model envisions to tackle more complex GPU MM scheduling scenarios when running a full-version of autonomous framework (e.g. Baidu Apollo) on next-generation platforms with more co-run tasks and larger design spaces, \textbf{\textit{which could not be easily solved by trial-and-error solutions}}.
\vspace{-2mm}
\subsection{Co-Optimizing Performance and Memory Footprint under Multitasking}
As was observed in Sec.~\ref{sec:mem_pres}, memory usage is a key factor in our target platforms (integrated CPU/GPU embedded systems). For the purpose of memory footprint reduction, a clear intuition is to change as many D policy applications as possible to M or H. However, in Sec.~\ref{sec:k_co} we also observe that aggressively doing so may cause a degraded GPU utilization, and prolonged GPU time of applications whose best memory management policy is identified to be D.

To achieve the goal of minimizing memory footprint while not sacrificing performance, in this section we establish several guidelines that are inspired by the opportunity disclosed in Fig.~\ref{fig:interleave} to determine GPU MM assignments for applications. For incoming tasks assigned with default D policy, our guidelines will determine if needs to further switch their memory management policies to the M or H policy. These guidelines establish criteria to determine whether the latency penalty and the memory reduction benefits are sufficient for any such switch to happen.
We then propose a runtime scheduler that strategically co-runs kernels to recover any degradation of performance resulted from the previous step. 

\noindent\textbf{Memory Footprint Minimization.} 
Using the performance model proposed in Sec.~\ref{sec:ana_model}, we calculate the values of $O_D$, $O_M$, and $O_H$ for each application. The best performance-oriented memory management policy for that application can then be identified. Next, pertinent to our design goal to minimize memory footprint, we establish guidelines which make sure that only specific types of D policy applications can be converted to M or H (i.e., any such conversion should not incur excessive performance loss), and shall prevent too many D policy kernels to be converted to M or H in order to leave enough room for multitasking optimization which requires overlapping a D policy kernel with an M or H policy kernel. Also recall that too many M/M co-run is not preferred as we explained in Sec.~\ref{sec:k_co}.

%%%%%%%%%%%
%figure 5 original location
 \begin{figure}[t]
\centering
\includegraphics[width=\linewidth]{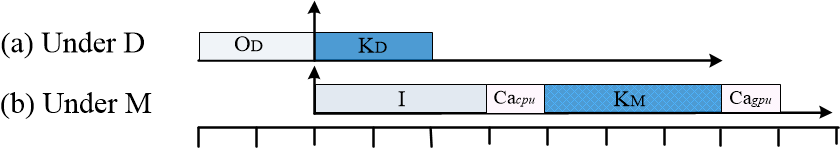}
\vspace{-6mm}
\caption{Illustration of kernel execution under D and M.
{\footnotesize{
The kernel execution $K_D$ is overlapped with idle period I of $K_M$.
}}
}
 \label{fig:example}
\vspace{-6mm}
 \end{figure}

Before describing our established guidelines, we present a motivational example.
Fig.~\ref{fig:example} shows an example scenario of an application containing a kernel $K$ under the D policy and the M policy. In the scenario of Fig.~\ref{fig:example}(a), the D policy would incur data copy overhead of $O_D$, as discussed in Sec.~\ref{sec:ana_model}. Fig.~\ref{fig:example}(b) shows the same kernel under the M policy. As described in Sec.~\ref{sec:ana_model}, there is an idle period denoted by $I$,which is calculated as $S_{i}*l_{mapping}$ in Equ.~\ref{Tl} , then an overhead at kernel start time denoted by $Ca_{cpu}$, and an overhead at kernel completion denoted by $Ca_{gpu}$. This change from D to M only makes sense if this introduced overhead is sufficiently small. Moreover, Fig.~\ref{fig:example}(b) also depicts another crucial fact: of all the overheads introduced by the M policy (i.e., $I$, $Ca_{cpu}$, and $Ca_{gpu}$), only part of it can be utilized for overlapping, i.e., to overlap another kernel execution with $I$. This is because $Ca_{cpu}$ and $Ca_{gpu}$ are cache management procedures that cannot be interrupted or overlapped because any such overlapping can cause unpredictable latency~\cite{nvidia2014toolkit}. This kernel, if under the M policy, can benefit from an overlapping co-execution with another D policy kernel during its idle period.

\noindent\textbf{Guideline 1:} \textbf{\textit{As we discussed in the previous example, the change of policy from D to M or H for an application should only happen if the loss of overhead is sufficiently small.}} This loss of overhead can be calculated as $O_M - O_D$ for M and $O_H - O_D$ for H.  However, imposing a limit on how much extra overhead is tolerable is not trivial. As we clarified in our example of Fig.~\ref{fig:example}, a loss in overhead can be acceptable because of the potential to overlap another kernel. However, if the idle portion of the overhead under M or H is too small compared to the total overhead, the probability of finding another kernel that would fit in that portion would become minuscule. Moreover, the recoverable overhead will be so small that the overall effect of overlapping co-execution would become negligible. Thus, we define the effective recoverable portion of the overhead due to conversion to M or H to be $I - Ca_{cpu} - Ca_{gpu}$ or $I-T_{c}$ respectively. A greater value of this parameter would encourage a conversion from D to M or H.  

We thus combine these requirements to present the following constraints:
%\vspace{-2mm}
\begin{equation}\vspace{-2mm}
O_M-O_D \leq I - Ca_{cpu} - Ca_{gpu},
\end{equation}
for M, and 
%\vspace{-2mm}
\begin{equation}
%\vspace{-2mm}
O_H-O_D \leq I - T_{c},
\end{equation}
for H. If the above equations are satisfied, the smaller of $O_M$ (M) or $O_H$ (H) should be selected. The intuition behind the above constraints is clear: we will convert a D policy application to an M or H policy application, if the effective recoverable portion of the overhead due to potential overlap (i.e., $I - Ca_{cpu} - Ca_{gpu}$) is at least the loss of overhead due to conversion (i.e., $O_M - O_D$).

\noindent\textbf{Guideline 2:} 
While Guideline 1 is crucial to prevent unnecessary performance loss, it is not sufficient to guarantee that there will be enough D policy kernels for the overlapping purpose. Thus, we establish Guideline 2 to ensure that there is always enough D policy kernels in the system to overlap with M and H policy kernels:
%\vspace{-1mm}
\begin{equation}
%\vspace{-1mm}
\label{equ:equal}
\sum\nolimits (N_M + N_H)\leq \sum\nolimits N_D ,
\end{equation}
in which $\sum\nolimits N_M+N_H$  is the total number of M and H policy kernels and $\sum\nolimits N_D$ is the number of D policy kernels.  Equ.~\ref{equ:equal} shows that \textit{\textbf{the number of D kernels shall be always more than the combined number of M and H kernels to guarantee that every overhead of M or H policy kernels can be recovered via an overlap with a D kernel}}.
Thus, for any D policy application, if it satisfies both constraints presented in the two guidelines, it will be converted into an M or H policy application for reduced memory usage.

\begin{figure}[t]
\centering
\includegraphics[width=\linewidth]{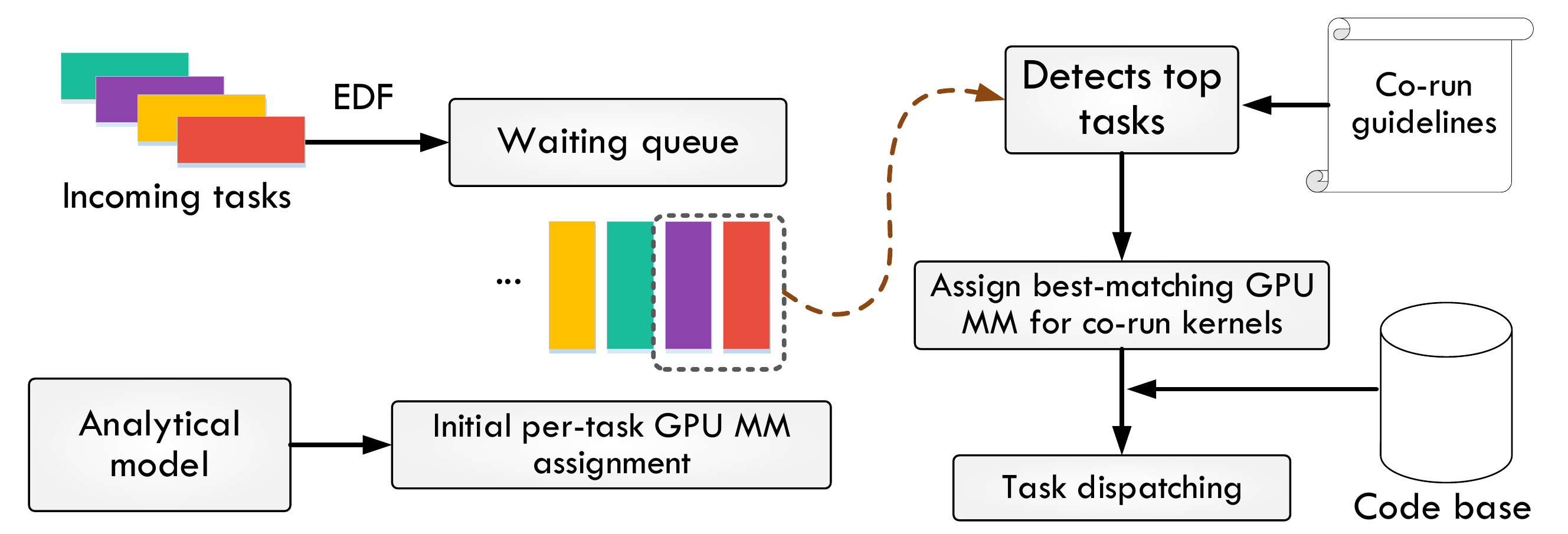}
\vspace{-7mm}
\caption{High-level overview of runtime scheduler.}
\label{fig:scheduling}
\vspace{-5mm}
\end{figure}

\noindent\textbf{Multitasking Scheduler.} We now explore the runtime overlapping opportunity under multitasking through overlapping the idle period of kernels that belong to the M/H policy kernels with the execution time of D policy kernels. To utilize this overlapping co-execution potential, we implement a runtime scheduler shown in Fig.~\ref{fig:scheduling}. The scheduler adopts the earliest deadline first (EDF) to maintain a waiting queue
%\footnote{Our design is compatible with any scheduling algorithm because the order of the waiting queue is not a factor in our design},
\footnote{The scheduling algorithm only impacts the sequence of different tasks in the waiting queue in Fig. 6 (i.e., before we apply the analytical model). Even though different scheduling algorithms lead to different tasks sequences, we can detect different top tasks and correspondingly assign the best-matching GPU MM for them and then dispatch one or two overlapped kernels to the GPU execution units. The key fact is that, if the co-running kernels can overlap with each other, the system performance can be enhanced (i.e., the latency is reduced). Therefore, our design optimization policy is compatible with different scheduling algorithms.} 
where the deadline of each kernel is set to its execution time. The runtime scheduler decides the best-matching GPU MM policy for newly arrived kernels and decides on which pair of kernels in the queue shall co-execute to benefit from overlapping. The scheduler checks each pair of kernels (suppose $K_1$ and $K_2$) in order, using the following criteria while meeting the guidelines in the meantime:
%\vspace{-2mm}
\begin{enumerate}[(1),topsep=1ex,itemsep=-1ex,partopsep=1ex,parsep=1ex]
    \item One kernel (suppose $K_1$) is executing under the D (or M) policy and the other is executing under the M (or D) policy (suppose $K_2$).
    \item The execution part of $K_1$ (or $K_2$) can fit in the idle period of $K_2$ (or $K_1$).
\end{enumerate}
%\vspace{-2mm}

\noindent For each round of kernel release, the scheduler checks whether a pair can be found for the kernel at the head of the queue. If so, the scheduler releases and co-executes both kernels together; otherwise, the scheduler only releases this single kernel.

In detail, specific code versions for each MM assignment are implemented by calling the specific MM APIs (e.g. \textit{cudaMallocManaged()}) in the application code. Then, the runtime scheduler is implemented using ZeroMQ library~\cite{zeromq} to assign a certain policy (D/M/H) to each incoming task %(e.g. YOLO, clustering), 
of which may consist of multiple kernels. Finally, the corresponding code of a task will be executed in the runtime when the task is dispatched by the scheduler. Note that no online profiling is needed during scheduling. We use an example to show the flexibility of our dynamic scheduling. During the runtime, a solo-run task YOLO\_A will use D policy according to the analytical model. However, if the scheduler detects it's followed by another task YOLO\_B, the scheduler will co-run them using M and D. If YOLO\_B is cancelled before YOLO\_A is dispatched, YOLO\_A will still use D policy. This shows that even the same task may be assigned to different MM policies at runtime according to its neighbor in the queue. Note that our approach can be generally applied to other workloads running on iGPU platforms.

\begin{figure*}[t]
\centering
\begin{minipage}[t]{.19\textwidth}
\includegraphics[width=1.1\columnwidth]{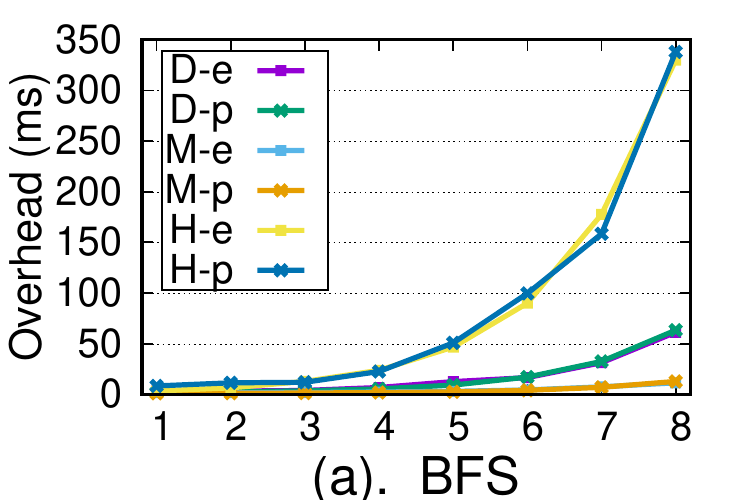}
\end{minipage}
\begin{minipage}[t]{.19\textwidth}
%\subfigure[]{
\centering
\includegraphics[width=1.1\columnwidth]{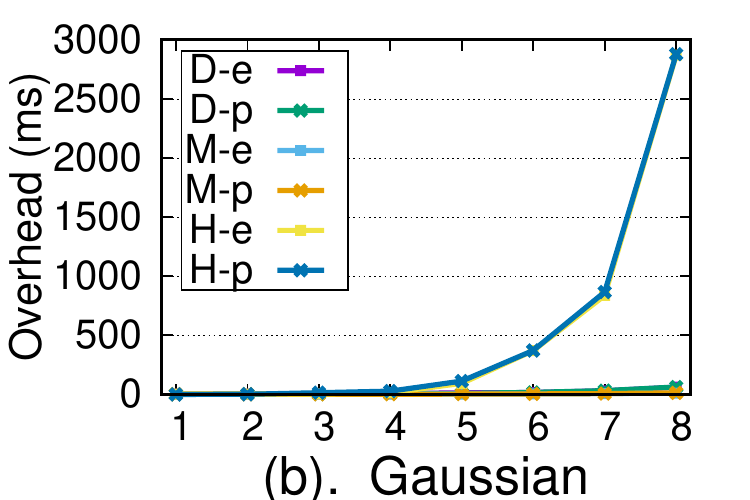}
%\caption{(b)}

\end{minipage}
\begin{minipage}[t]{.19\textwidth}
%\subfigure[]{
\centering
\includegraphics[width=1.1\columnwidth]{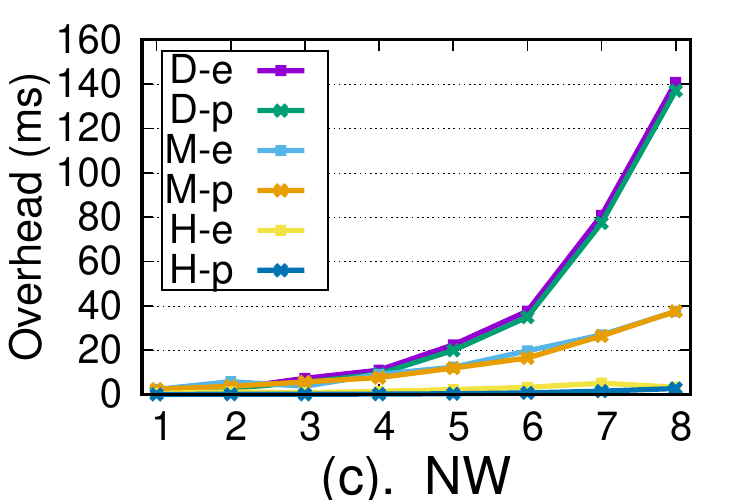}
%\caption{(c)

\end{minipage}
\begin{minipage}[t]{.19\textwidth}
\centering
\includegraphics[width=1.1\columnwidth]{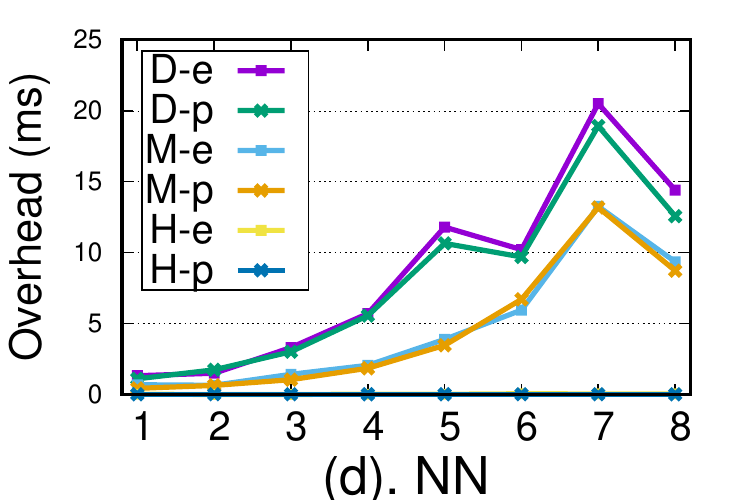}
%\caption{(d)}
\end{minipage}  
\begin{minipage}[t]{.19\textwidth}
\centering
\includegraphics[width=1.1\columnwidth]{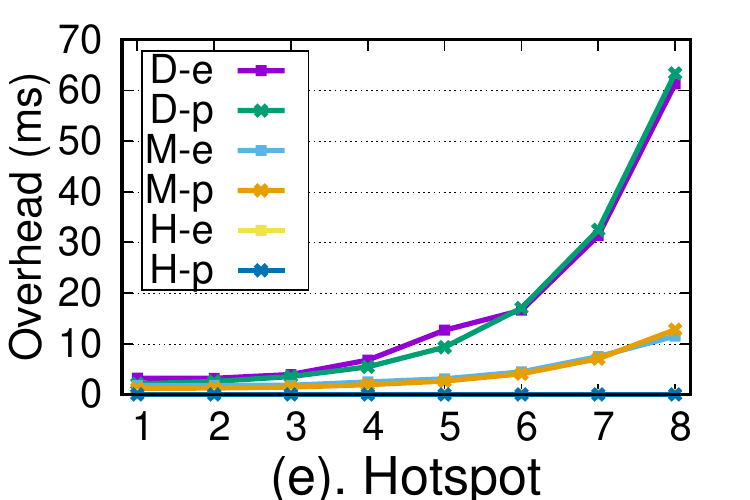}
%\caption{(e)}
\end{minipage}
\begin{minipage}[t]{.19\textwidth}
\centering
\includegraphics[width=1.1\columnwidth]{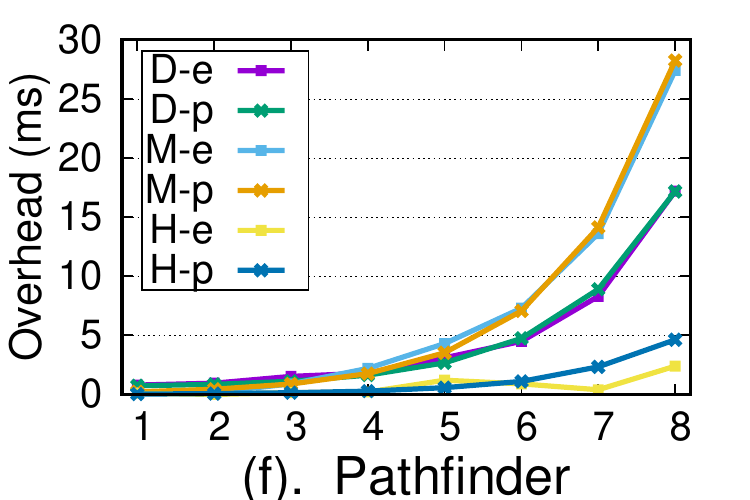}
\end{minipage}
\begin{minipage}[t]{.19\textwidth}
\centering
\includegraphics[width=\columnwidth]{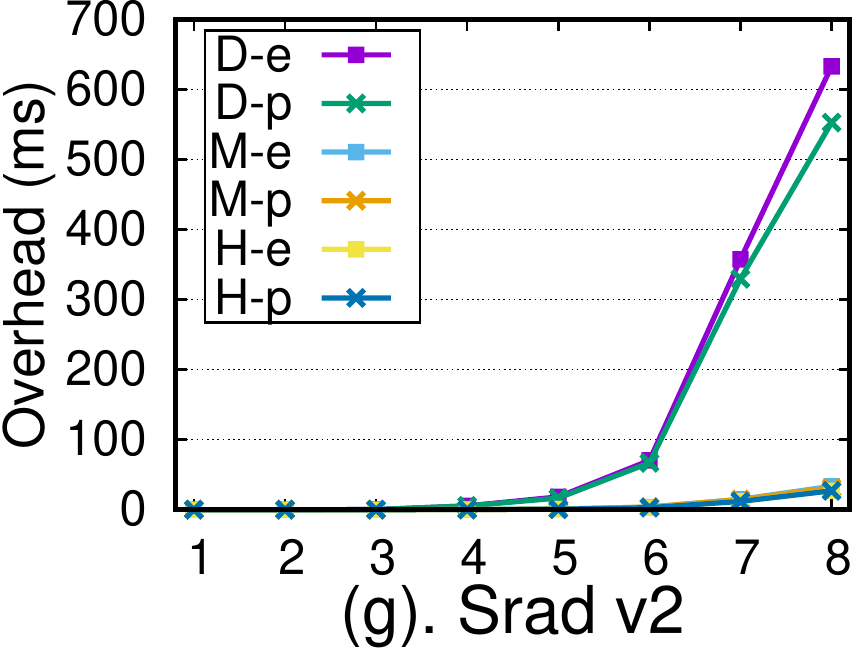}
%\caption{(e)}
\end{minipage}
\begin{minipage}[t]{.19\textwidth}
\centering
\includegraphics[width=\columnwidth]{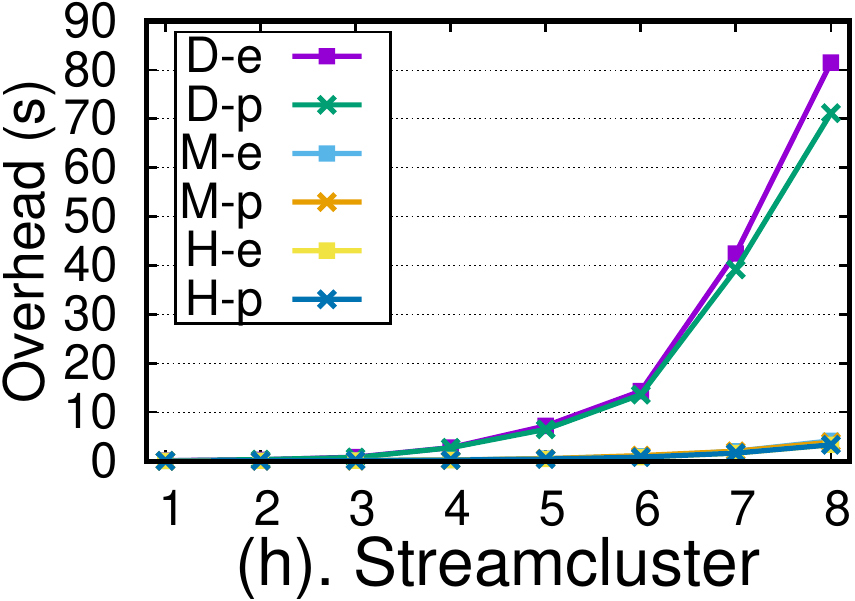}
%\caption{(e)}
\end{minipage}
\begin{minipage}[t]{.19\textwidth}
\centering
\includegraphics[width=\columnwidth]{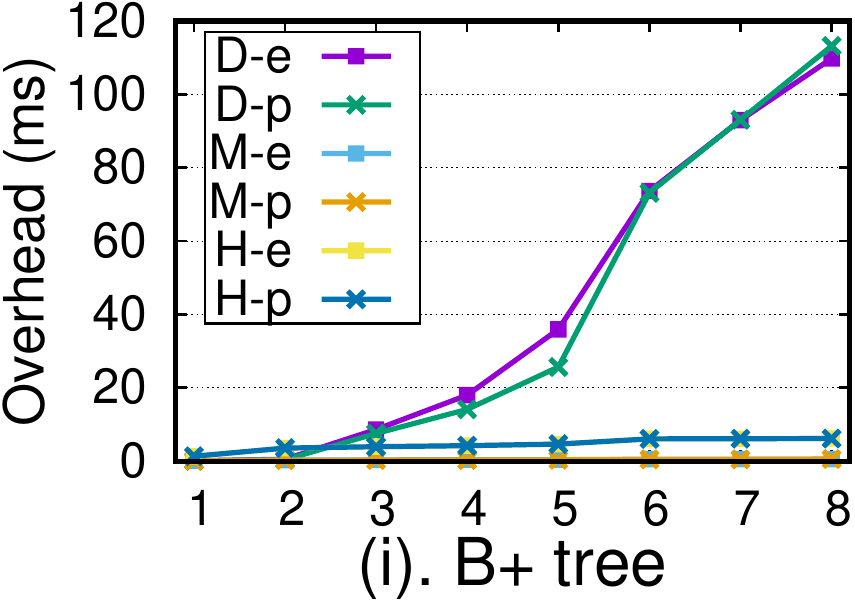}
%\caption{(e)}
\end{minipage}
\begin{minipage}[t]{.19\textwidth}
\centering
\includegraphics[width=\columnwidth]{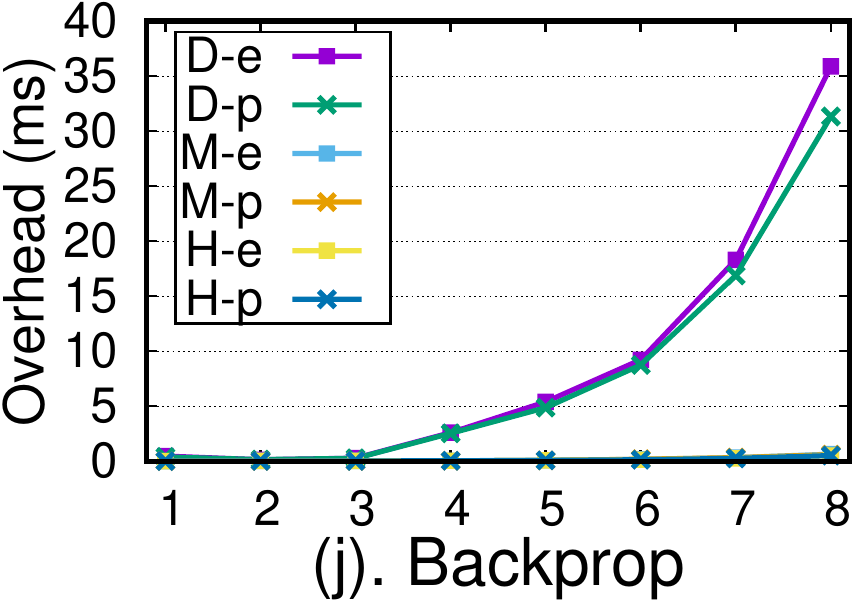}
%\caption{(e)}
\end{minipage}

\begin{minipage}[t]{.19\textwidth}
\centering
\includegraphics[width=1.1\columnwidth]{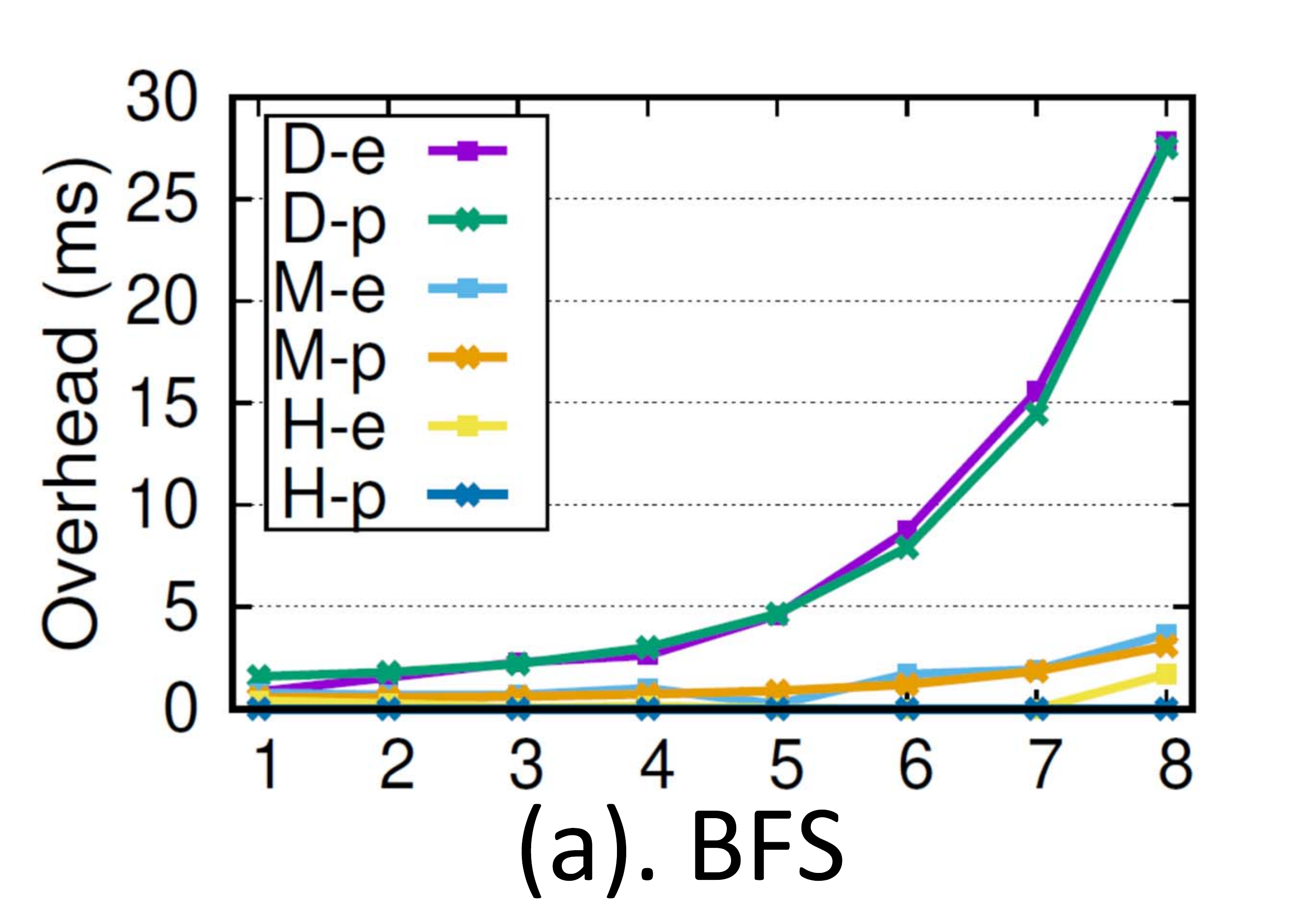}
\end{minipage}
\begin{minipage}[t]{.19\textwidth}
\centering
\includegraphics[width=1.1\columnwidth]{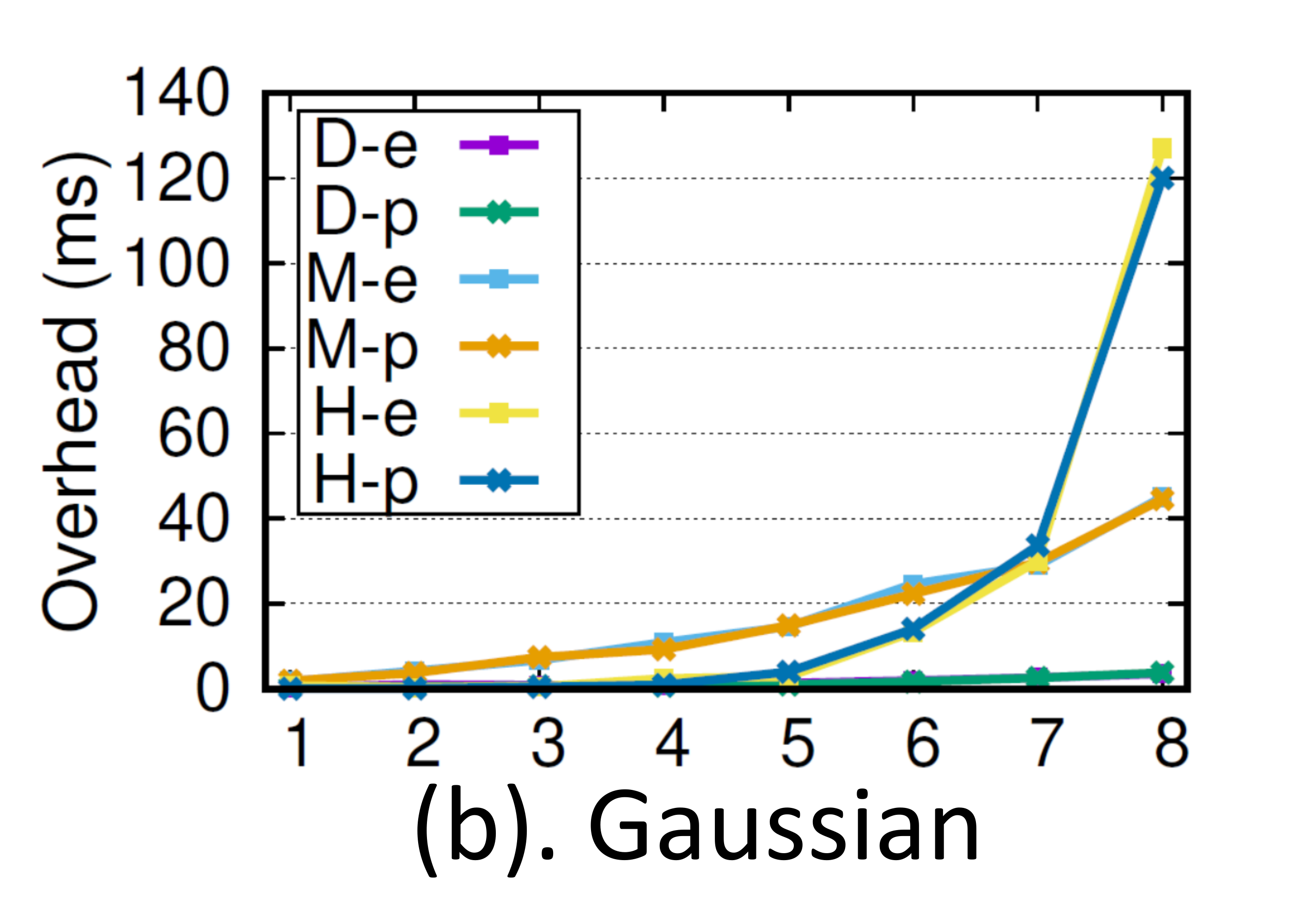}
\end{minipage}
\begin{minipage}[t]{.19\textwidth}
\centering
\includegraphics[width=1.1\columnwidth]{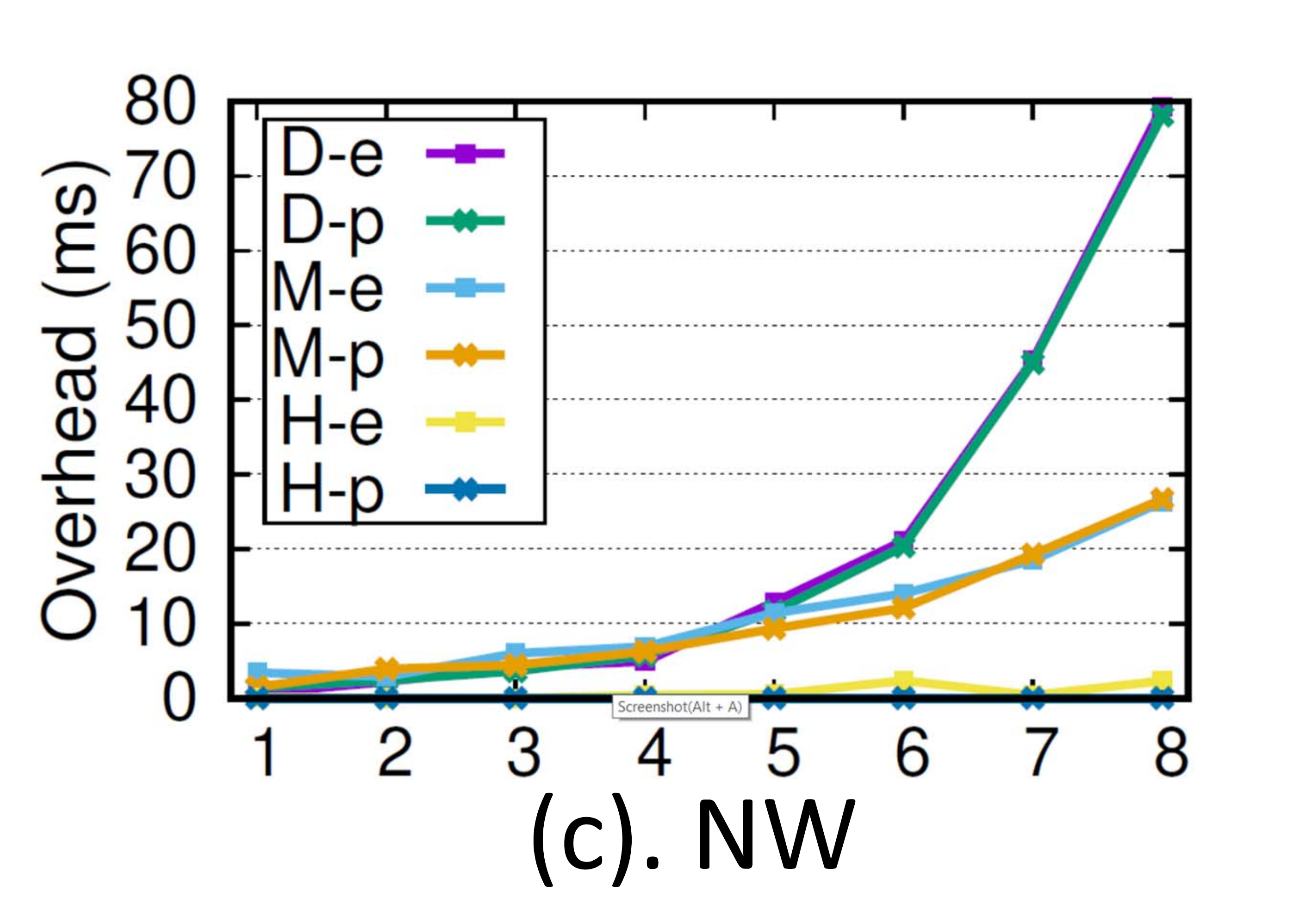}
\end{minipage}
\begin{minipage}[t]{.19\textwidth}
\centering
\includegraphics[width=1.1\columnwidth]{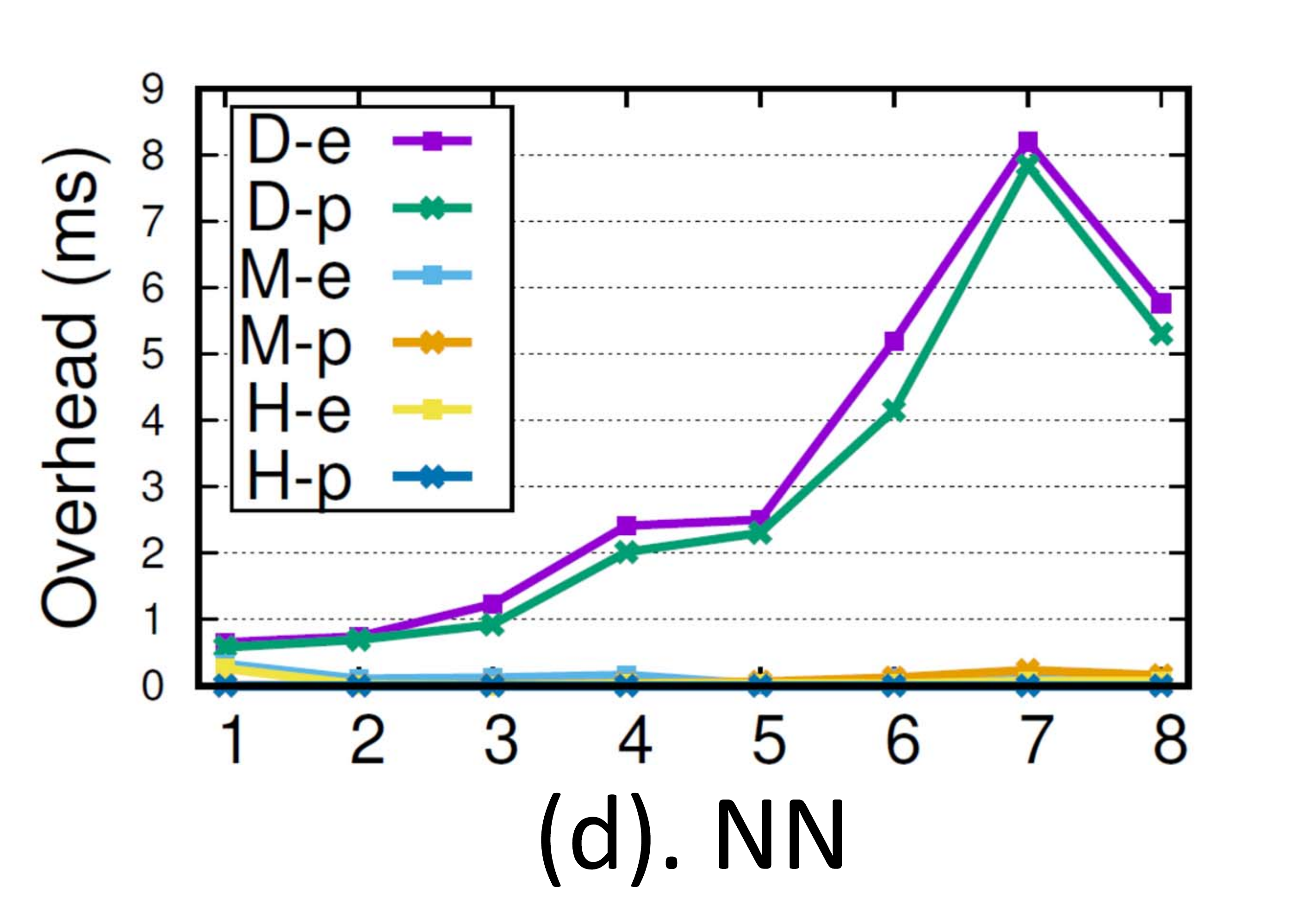}
\end{minipage}  
\begin{minipage}[t]{.19\textwidth}
\centering
\includegraphics[width=1.1\columnwidth]{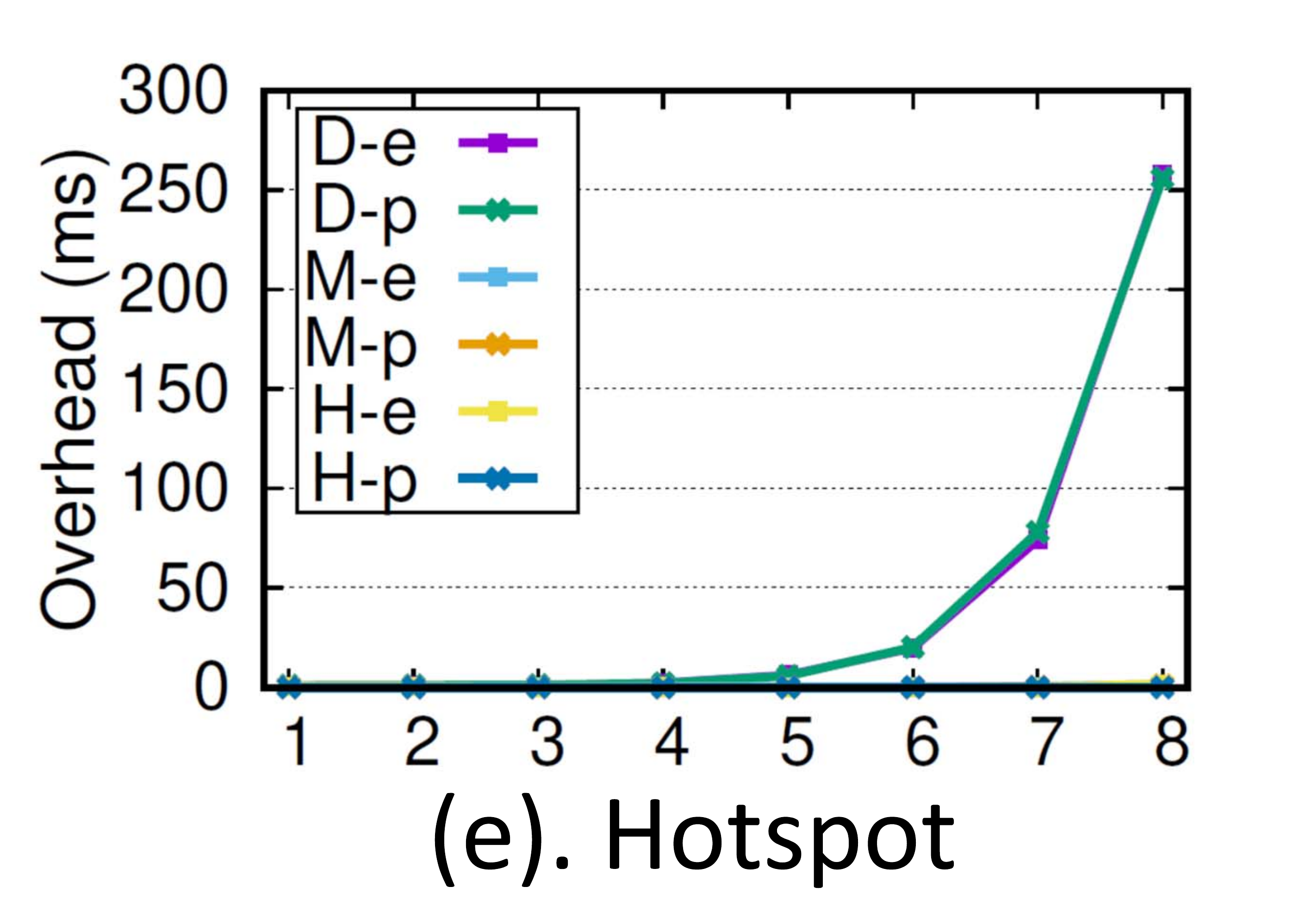}
\end{minipage}
\begin{minipage}[t]{.19\textwidth}
\centering
\includegraphics[width=1.1\columnwidth]{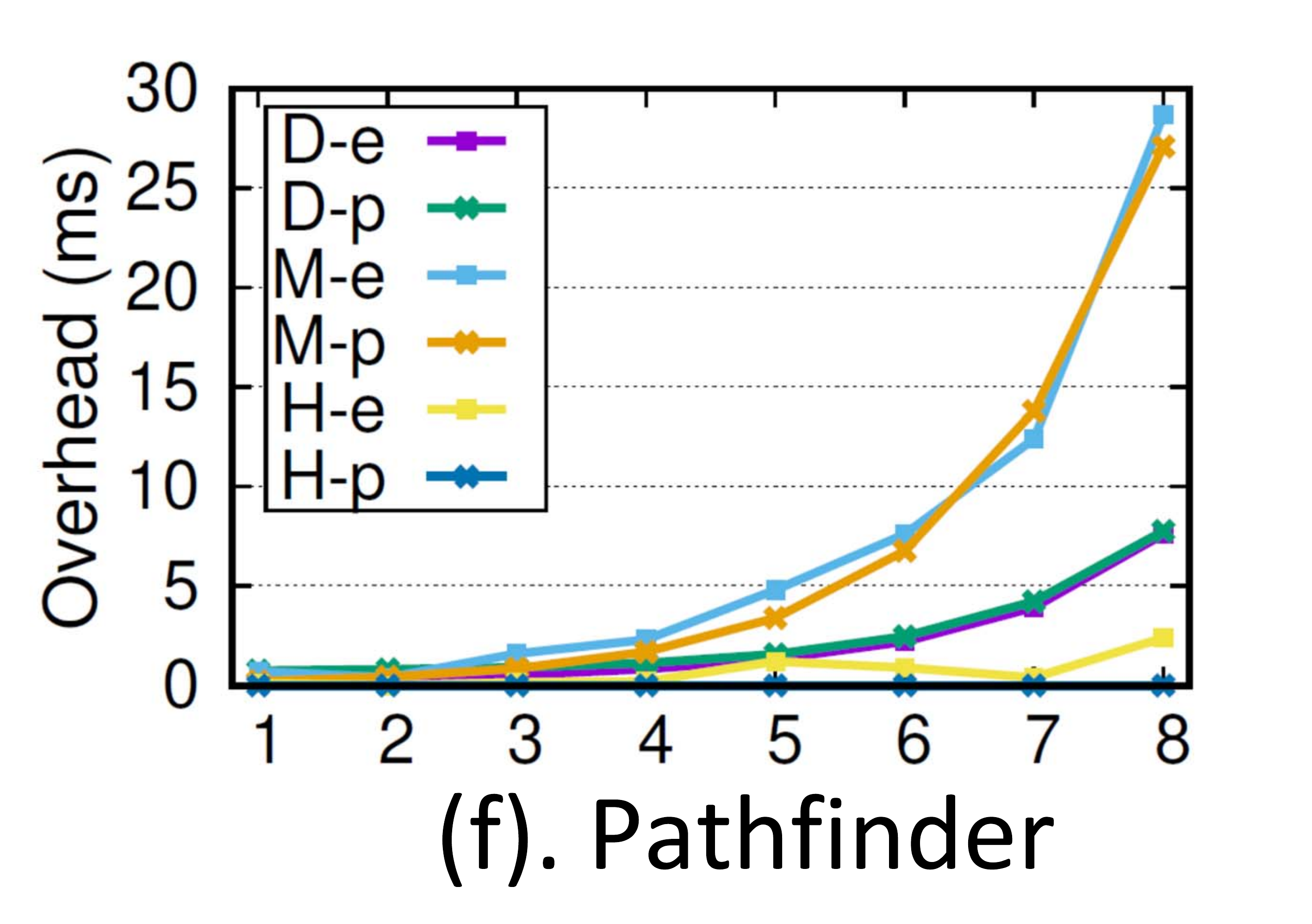}
\end{minipage}
\begin{minipage}[t]{.19\textwidth}
\centering
\includegraphics[width=\columnwidth]{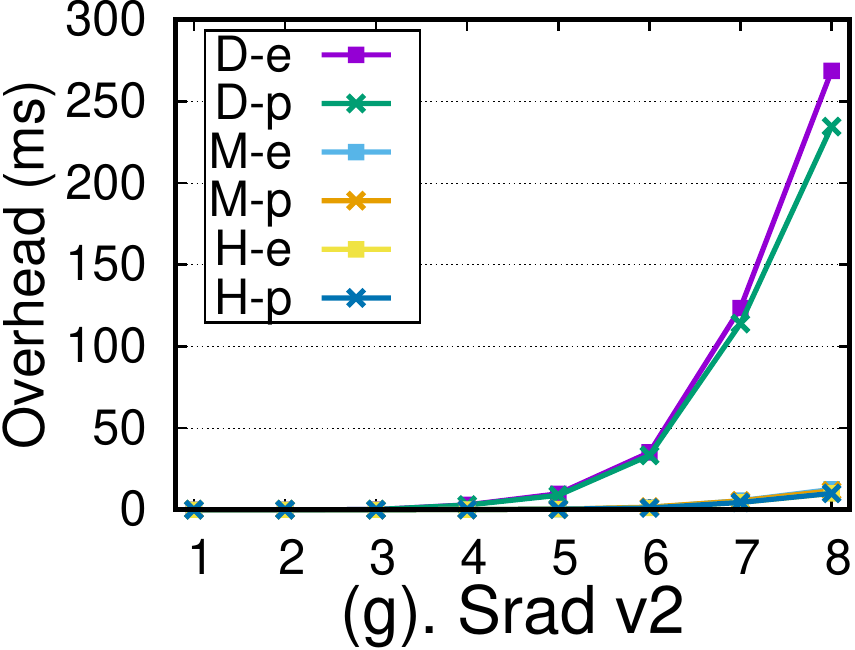}
%\caption{(e)}
\end{minipage}
\begin{minipage}[t]{.19\textwidth}
\centering
\includegraphics[width=\columnwidth]{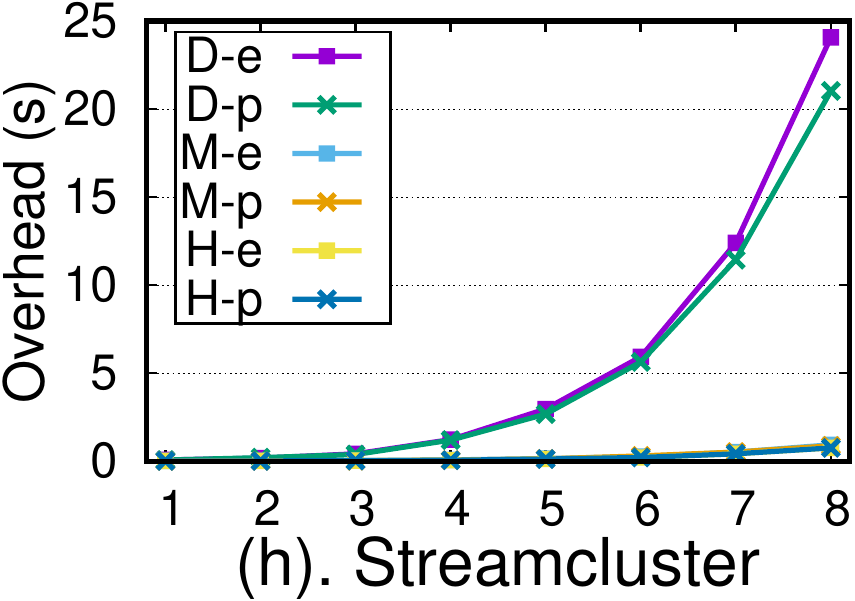}
%\caption{(e)}
\end{minipage}
\begin{minipage}[t]{.19\textwidth}
\centering
\includegraphics[width=\columnwidth]{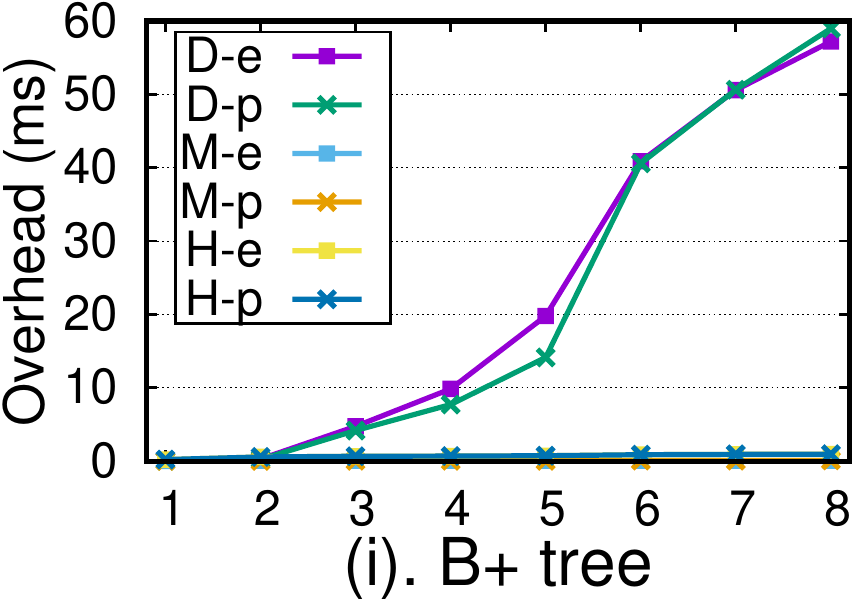}
%\caption{(e)}
\end{minipage}
\begin{minipage}[t]{.19\textwidth}
\centering
\includegraphics[width=\columnwidth]{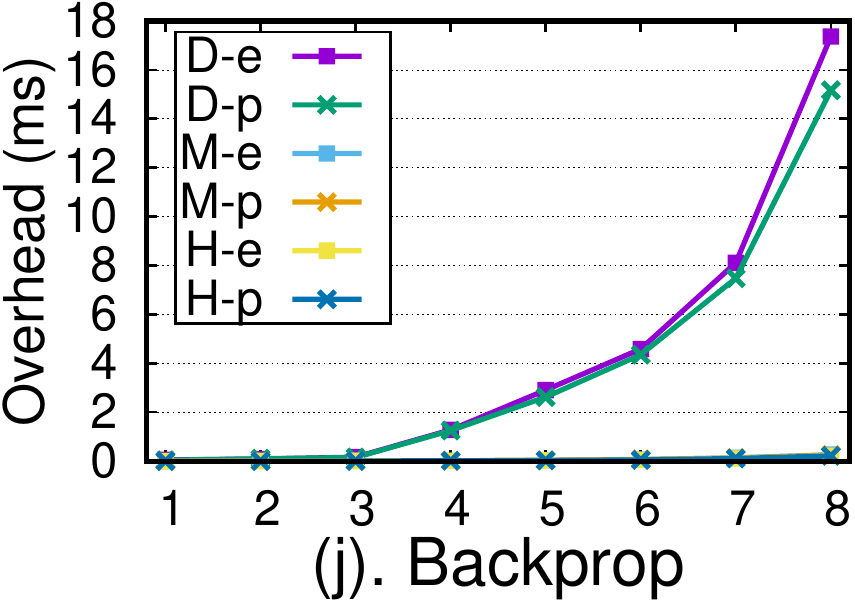}
%\caption{(e)}
\end{minipage}
\caption{Model validation on Jetson TX2 (row 1 and row 2) and Xavier AGX (row 3 and row 4). 
{\footnotesize{We compare both experimental measurements (D-e/M-e/H-e) and the predicted values (D-p/M-p/H-p) calculated from our performance model. We calculate the geometric error of each benchmark under the 8 input size. The largest error indicates the worst case of the model on the type of platform. The geometric error of the ten benchmarks indicates the average error of the model on the type of platform. 
}}
}
\vspace{-5mm}
\label{fig:Model}
\end{figure*} 

\vspace{-1mm}
\section{Evaluation}
\vspace{-1mm}

\subsection{Experimental Setup}
%\vspace{-1mm}
\noindent\textbf{Test Workloads and Methodology.}
We evaluate our design with both well-established micro-benchmarks and real-world autonomous device applications. First, we validate our analytical model using micro-benchmarks from the Rodinia benchmark Suit ~\cite{rodinia} because it includes applications and kernels targeting multi-core CPU and GPU platforms and covers a diverse range of application domains. Each program in the Rodinia benchmark exhibits various types of parallelism, data access patterns, and data-sharing characteristics as well as synchronization techniques. Nonetheless, these benchmarks can be roughly categorized into two types: kernel-intensive and memory-intensive. Considering that we focuses on the memory behavior of applications, we use all the memory-intensive and two kernel-intensive benchmarks to evaluate our analytical model. Second, we evaluate our co-optimization scheduling policy. According to ~\cite{landaverde2014investigation}, the memory-intensive benchmarks can be further categorized into the HtoD-memory-intensive and Balanced-memory-intensive categories. The HtoD-memory-intensive indicates that HtoD transfer is significantly more time-consuming than the DtoH transfer, while the Balanced-memory-intensive indicates that the HtoD and the DtoH transfers are comparable in terms of latency. Therefore, we select two representative benchmarks and co-execute them with the other two kernel-intensive benchmarks to evaluate our co-optimization scheduling policy. Finally, we evaluate our full design using two real-world GPU-intensive embedded autonomous systems: the object detection component of DJI drone software and the perception module of Autoware~\cite{autoware}.

\noindent\textbf{Hardware Platforms.}
The model validation and GPU MM method evaluation are implemented on both Parker and Xavier platforms. We use NVIDIA TX2 and AGX as the representative for Parker and Xavier SoCs. These platforms are widely adopted by many drone companies, such as DJI and IFM ~\cite{djidrone2017sensing,ifm} and used in prototype design and product performance evaluation. Recently, the Parker-based TX2 and Drive PX2 are introduced to enable many driverless car applications. Autoware~\cite{autoware}, the system software for driverless cars, is well-supported on both platforms. Autoware + Drive PX2 AutoChauffeur is used to represent a very realistic case study of autonomous vehicles. Consequently, in our case studies, we evaluate the memory usage and timing performance as well as GPU utilization of the object detection component of DJI drone software and the Autoware perception modules under three configurations. 

Table~\ref{tbl:workloads} shows the evaluated application (column 2), the corresponding tested platform (column 3), the MM policy chosen by our approach (column 4), and the number of kernels contained in each application (column 5). Note that there are three MM policies chosen for Autoware as it contains three separate modules (detailed in Sec. 5.3).
%\vspace{-1mm}
\subsection{Evaluation on Microbenchmarks}
%\vspace{-1mm}
\noindent\textbf{Validation of the Analytical Performance Model.}
We run all six benchmarks on both TX2 and AGX under three GPU MM policies. We compare both experimental measurements (D/M/H-e) and the predicted values (D/M/H-p) calculated from our performance model, as shown in Fig.~\ref{fig:Model}. For each benchmark, we test 8 different input data size (e.g., in Pathfinder, the columns of matrix for computation ranges from 160, 320, ... to 20480). We can observe that the measured time of all the benchmarks in experiments well matches their performance calculated by our model. Thus, our model proves to be effective and accurate enough to predict the overhead of single application under a given GPU MM policy. To examine how well the model works on TX2/AGX platform, we provide the errors of the model as follows: on TX2, in worst case, the geometric mean of the error is 14.9\% for D policy, 16.4\% for M policy and 23.1 \% for H policy; in average case, the error is 9.6 \% for D, 11.8\% for M and 6.5\% for H policy. When we do validation on AGX, in worst case, the error is 15.0 \% for D policy, 24.3 \% for H policy and 14.3 \% for H policy; in average case, the error is 10.9\% for D, 14.7\% for M and 2.39\% for H policy.

\begin{table}[t]
\centering
\small{
%\resizebox{\columnwidth}{20mm}{
\begin{tabular}{@{}lllll@{}}
\toprule
Catagory          & Workload    & Testbeds  &MM   &kernels  \\\midrule
benchmark  &BFS   & TX2/AGX     &M/M     &24       \\ 
benchmark  &Pathfinder  & TX2/AGX     &M/D     &512         \\
%benchmark  &B+tree  & TX2/AGX     &M/M     &2      \\
benchmark  &NW	        & TX2/AGX     &M/D     &320      \\ 
benchmark  &NN          & TX2/AGX     &M/M     &1       \\ 
%benchmark  &Backprop  & TX2/AGX     &M/M     &2      &balanced-memory-intensive   \\ 
%benchmark  &Srad\_v2  & TX2/AGX     &M/M     &4      &balanced-memory-intensive   \\ 
%benchmark  &Streamcluster  & TX2/AGX     &M/D     &1790      &balanced-memory-intensive   \\ 
benchmark  &vect-add    & TX2/AGX     &M/M     &1  \\
benchmark  &Gaussian    & TX2/AGX	  &D/D     &1024     \\
benchmark  & Hotspot     & TX2/AGX     &M/M     &1     \\\midrule
Real App.  &DJI Drone   & TX2/AGX     &M/M     & 180        \\
Real App.  &Autoware    &PX2          &{D,M,M}      & 599   \\ \bottomrule
\end{tabular}}
%}
\vspace{2mm}
\caption{Evaluated workloads and hardware platforms.}
\vspace{-10mm}
\label{tbl:workloads}
\end{table}

\begin{figure}[t]
\vspace{-2mm}
\centering
\includegraphics[width=1.0\linewidth]{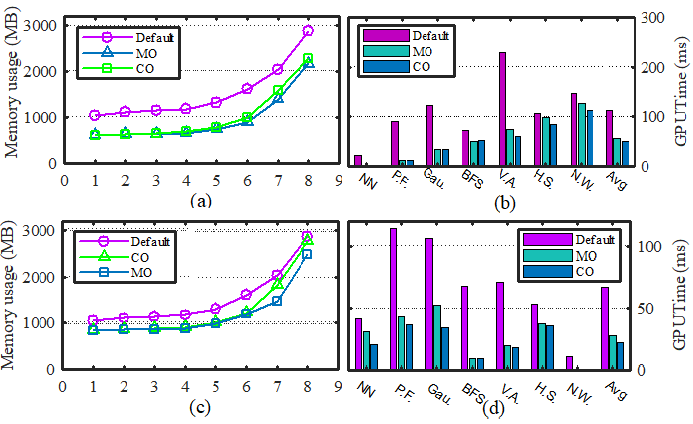}
\vspace{-8mm}
\caption{Memory usage and GPU time of Benchmarks.
{\footnotesize{
The x-axis represents the 8 different input data sizes. The y-axis shows the worst-case memory usage of all benchmarks corresponding to a specific input data size.
}}
}
\vspace{-2mm}
\label{fig:AGXbenchmark}
\end{figure}

\noindent\textbf{Memory Usage/Performance Co-optimization.}
We then evaluate the effectiveness of our co-optimization scheduler. We co-execute all seven applications with a specific kernel number for each application, as shown in Fig.~\ref{fig:AGXbenchmark}. We apply three configurations for the system GPU MM setup. Namely, the Default (D), Memory Optimized (MO), and Co-Optimized (CO). D is the baseline setup where only the traditional device allocation policy is assigned to each kernel. MO configuration only adopts Managed Memory policy to each kernel with the goal of minimizing the memory footprint at all cost. In CO, all kernels will be scheduled by our memory footprint/performance co-optimization scheduler with smart policy switching and kernel overlapping. We report the results of the memory usage and GPU time under three GPU MM setups (D, MO, CO) and two hardware configurations (TX2 and AGX).

Fig.~\ref{fig:AGXbenchmark}(a) shows the memory usage of the seven benchmarks co-executing under different configurations on NVIDIA Jetson TX2. Generally speaking, the memory usage will increase as the input data size increases. Under the configuration D, the memory usage reaches the maximum while the memory usage drops to the minimum under configuration MO. This result is reasonable since the default configuration would consume extra memory space compared to MO. We observe that, under CO configuration, only Gaussian doesn't change its GPU MM policy and still maintains D due to guideline 2 listed in our design, while all the other benchmarks adopt M as their policy. As a result, the memory usage almost has no change from MO to CO. Meanwhile, the overall timing performance of all benchmarks benefit from the fact that Gaussian adopts D while the others adopt M under CO, as is shown in Fig.~\ref{fig:AGXbenchmark}(b), where the x-axis represents the seven benchmarks and the average case and y-axis represents the corresponding GPU time. We observe that the response time of Gaussian under CO is 75\% less than default, and doesn't increase compared to MO. Take all benchmarks and the average case into consideration, the improvement of response time of configuration CO ranges from 21.2\% to 92.2\% compared to default, and ranges from -5.5\% to 18.7\% compared to MO; on average, the improvement of CO is 54.6\% compared to default, and 10.0\% to MO.   

\begin{table}[t]
\centering
\resizebox{\columnwidth}{10mm}{
\begin{tabular}{@{}llllll@{}}
\toprule
        & Ro.(TX2)  & Ro.(AGX) & UAV(TX2)  & UAV(AGX)  & Car(PX2)   \\ \midrule
D        & 74.6\%     & 63.1\% &  70.6\%  & 26.2\%   & 84.3\% \\
MO        & 56.2\%     &  57.3\% &  28.3\%  & 22.4\%   & 72.9\% \\
CO        & 74.6\%     &  63.0\% &  29.1\%  & 23.3\%  &  80.7\% \\ \bottomrule
\end{tabular}
}
\vspace{0mm}
\caption{GPU utilization of running different benchmarks on different platforms. {\footnotesize{Ro. indicates benchmarks. UAV indicates Drone. Car indicates Autoware.}}}
\vspace{-6mm}
\label{tbl:evalutilization}
\end{table}

We also evaluate the optimization policy on the Xavier-based Jetson AGX platform to present that our approach can be flexible, future-proof, and architecture independent. The results are shown in Fig.~\ref{fig:AGXbenchmark}(c) and (d). For the Xavier platform, the memory usage and timing performance of benchmarks follow the same principle.
Where Xavier differs from Parker is that with increasing data size, the memory usage under CO configuration becomes noticeably larger than that under MO configuration. This is due to the fact that, Xavier is a high-level architecture and has 2.5x memory access speed and faster transfer speed than Parker, which makes H and D more beneficial. Thus, each benchmark may perform differently under the same GPU MM policy. Note that on Parker, only Gaussian adopts the D policy while the others all adopt M. However, on Xavier, NW, Pathfinder and Gaussian all adopt D. %as their own MM policy while the other still adopt M 
Therefore, with data size increasing, the benchmarks under CO consume more memory space considering that the D policy requires extra space allocation. For the latency performance, the overall GPU time of the average case on Xavier is much less than that of Parker, which is due to the higher computation power and efficiency of Xavier.

Besides, we measure the GPU utilization of the benchmarks co-execution under three different configurations, as shown in Table~\ref{tbl:evalutilization}. We can note that the GPU utilization is only 56\% under MO. While the GPU utilization reaches about 75\% under D and CO. From the analysis above, we can safely conclude that our approach provides a potential opportunity to minimize memory footprint while enhancing the latency performance of applications running on GPU and GPU utilization, particularly under multitasking scenarios. 
As a matter of fact, if an application is executed under the M policy, its memory footprint is definitely reduced compared to the footprint under the D policy, which is the key benefit of the unified memory. In the multitasking scenario, if all applications adopt the M policy (i.e., the MO configuration), the memory footprint can be guaranteed to be minimized compared to the default or the CO configuration. On the other hand, by using the CO configuration, the latency performance of the system can be significantly enhanced if there are enough D kernels and M kernels to overlap, although in this case the memory footprint may not be minimized compared to the MO configuration. Nonetheless, our CO configuration can co-optimize the system performance and memory footprint simultaneously, because under CO, some workloads will adopt the M policy to reduce the memory footprint and these M kernels can overlap with the D kernels to reduce system latency, as our evaluation also shows.

\vspace{-2mm}
\subsection{Case Studies}
%\vspace{-2mm}
\noindent\textbf{Case I: Object Detection for Drone Navigation.}
Obstacle detection is a key feature for safe drone navigation. Currently, DNNs are widely adopted by commercial drone products to achieve efficient and accurate object detection. One of the representative examples is YOLO~\cite{joseph2016yolo}, which is known for its high accuracy, reliability, and low latency. DJI, as a technologically-advanced drone manufacturer, mainly utilizes YOLO to implement its object detection~\cite{djidrone2017sensing}. 
\begin{figure}[t]
\vspace{-6mm}
\centering
\includegraphics[width=1.1\linewidth]{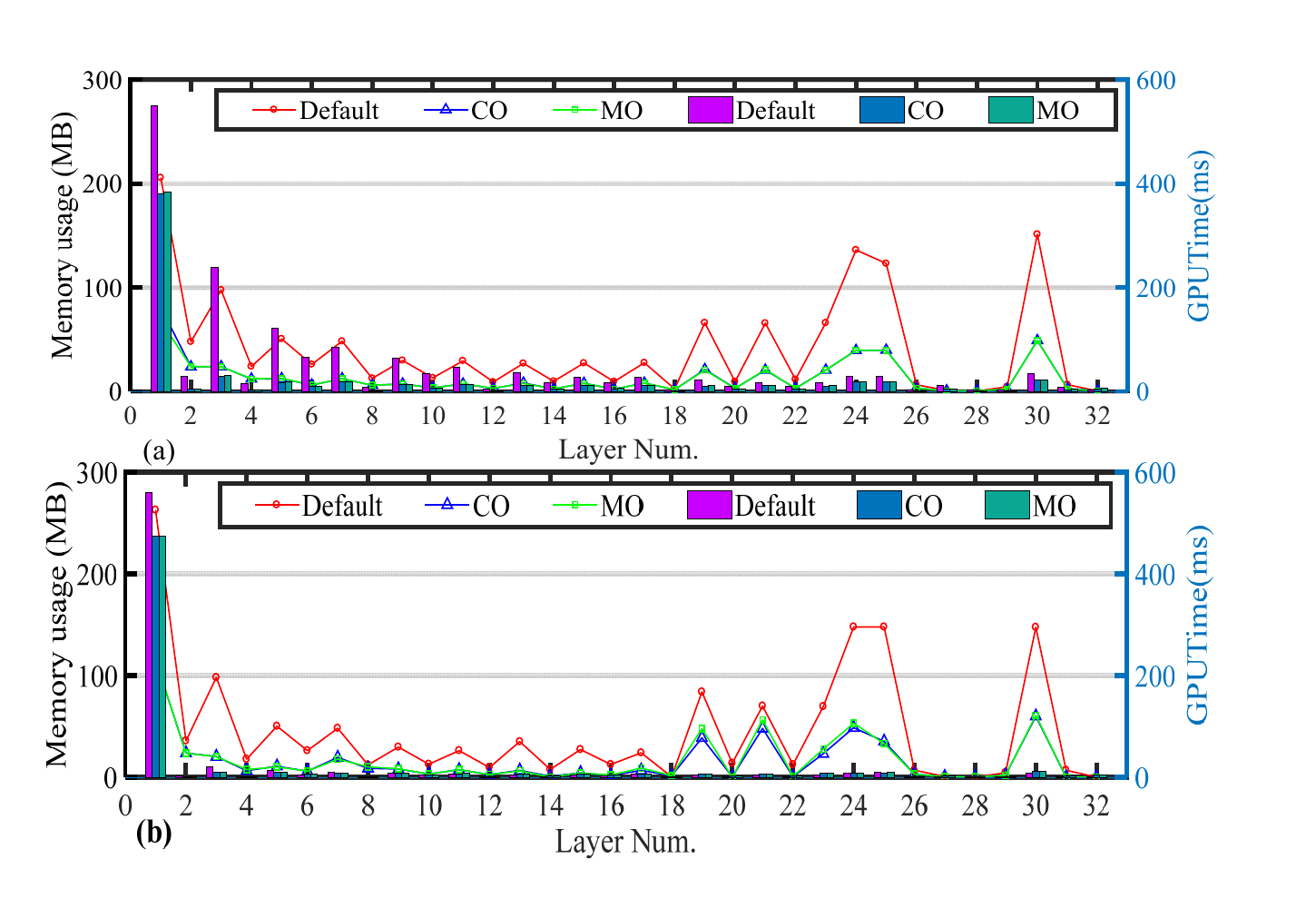}
\vspace{-12mm}
\caption{Per-layer memory usage and GPU time breakdown for the  drone-based case study}\label{droneperformanceonAGX}
\vspace{-2mm}
\end{figure}

\begin{figure}[t]
\centering
\vspace{-4mm}
\includegraphics[width=1.0\linewidth]{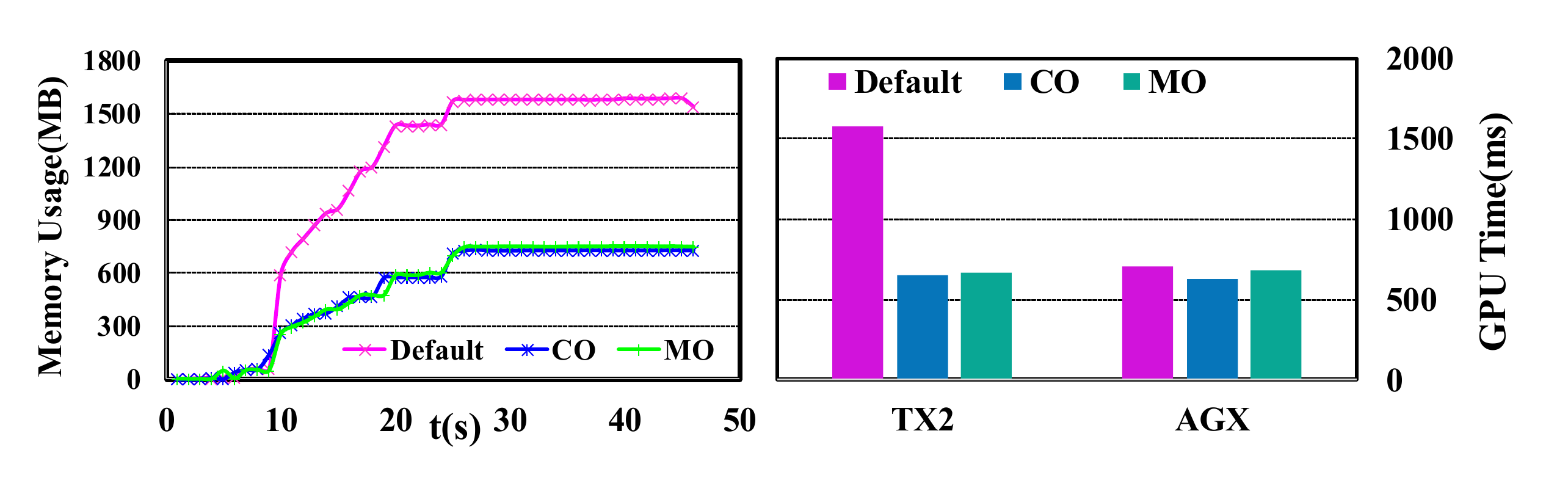}
\vspace{-10mm}
\caption{Drone performance on TX2 and AGX}\label{dronelatency}
\vspace{-5mm}
\end{figure}
However, as we analyzed in Section~\ref{sec:mem_pres}, YOLO typically causes memory pressure on integrated platforms. To that end, we apply our optimization policy to YOLO on the Parker-based Jetson TX2 to analyze its memory usage and timing performance. Fig.~\ref{droneperformanceonAGX}(a) shows the results, where the x-axis represents the layer number corresponding to a specific YOLO layer (YOLOV2.cfg), the y-axis on the left represents the memory consumed by each layer (line), and the y-axis on the right represents the GPU time taken by each layer (bar). Generally speaking, since convolution layer (e.g., layers 1, 3, and 5) involves more computation, it consumes more memory space and causes larger latency in computation than other types of layers, such as maxpool layer or route layer (corresponding to layers 2, 26, and 29). Moreover, since the first layer has to allocate enough space for the raw input data, it consumes extra memory space and requires more time compared to the subsequent layers due to kernel launch initialization. Clearly, we observe that under CO, the memory usage of each layer of YOLO is much less than the default configuration. Furthermore, by comparing the GPU time performance under different configurations, we find that the time consumed by each layer under CO is also much smaller than the default configuration. Besides, for YOLO, the MO configuration is similar to CO in the sense that each layer almost has the same memory usage and timing performance under the two configurations. Moreover, when we evaluate the memory usage and timing performance of YOLO on AGX, depicted in Fig.~\ref{droneperformanceonAGX}(b), we observe that the behavior of each layer is almost the same as the Parker platform. 

We also compare the overall memory usage and GPU time performance, as shown in Fig.~\ref{dronelatency}. Due to the space limit we only report the memory usage on TX2 since it is similar to the AGX, as shown in left-hand part of Fig.~\ref{dronelatency}. We can observe that the memory usage jumps fast with the increase of execution time under default configuration. While the memory usage of CO and MO configurations are significantly reduced. On the other hand, the Xavier SoC adopts the NVIDIA Volta GPU architecture, which is much more powerful than the Pascal architecture used in Parker. Therefore, under the default configuration, the response time of most layers is just a little worse than the time under the CO configuration. However, undeniably, even though the timing performance under CO doesn't improve much over default, the application consumes much less memory footprint, which is quite significant for our targeted memory-limited platforms. Besides, we measure the GPU utilization of executing drone application on TX2 and AGX, and find that, under D configuration, the GPU utilization is higher than MO and CO configurations, as shown in Table~\ref{tbl:evalutilization}, which follows the same trend of microbenchmark implementation on both platforms.

\noindent\textbf{Case II: Perception of Autoware.}
Autoware is an open source software that utilizes the Robot Operating System (ROS) as an underlying platform. Autoware is designed for autonomous driving, and effectively integrates a perception, planning, and control module to provide a complete autonomous-driving solution. The perception module provides information to subsequent planning and control modules, and typically consumes much memory space to be able to execute on the platform. Since all GPU-related sub-modules of Autoware are within the perception module, we focus on evaluating the memory usage and timing performance of the perception module in this case study, which includes two YOLO-based vision instances and one Euclidean clustering-based LIDAR instance. We implement Autoware on Drive PX2 and measure the memory usage and GPU time of three nodes when they receive one frame of input video data. Note that since Autoware is not supported on AGX/L4T31.1, we only test Autoware on PX2 to evaluate our optimization policy.

\begin{figure}[t]
\centering
\includegraphics[width=1.0\linewidth]{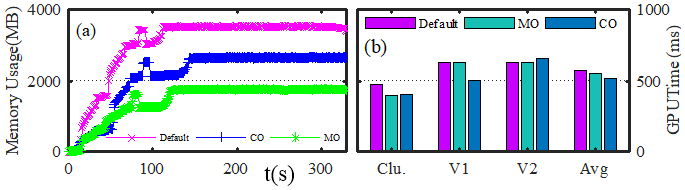}
\vspace{-8mm}
\caption{Autoware performance on Drive PX2 w.r.t. memory usage (a) and GPU time (b)}\label{autoware}
\vspace{-6mm}
\end{figure}

Fig.~\ref{autoware}(a) shows the total memory usage of the three nodes under different configurations on NVIDIA Drive PX2. The x-axis represents the time, while the y-axis represents the real-time memory utilization of the three nodes co-executing. As is evident in the figure, the memory allocated for all three nodes increases over time. Fig.~\ref{autoware}(a) shows that the memory consumption dramatically increases at the beginning. This is due to the fact that memory allocation is mainly implemented when the model of each node is loaded (e.g., loading the DNNs configuration for the vision node), and this procedure always happens prior to the node execution.  
After the three nodes are completely loaded on Autoware, the memory allocation completes and the memory usage change becomes imperceptible. As expected, the memory usage is minimal under MO, and has a reduction of 34 \% in memory usage compared to CO and 50\% compared to default configuration. Since Parker iGPU has deficient memory resources while the perception modules of Autoware have high requirements on memory space, by adopting CO or MO, the memory space can be saved a lot such that more functional nodes can be enabled simultaneously, such as localization, planning, etc.  

Furthermore, we measure the timing performance of the GPU-dependent three nodes. Fig. \ref{autoware}(b) shows the results, where the x-axis represents the specific nodes as well as their average, and the y-axis represents the GPU time. The GPU time under default is the baseline. We observe that the average GPU time of each node under MO is reduced by 6\% and the time under CO is reduced by 10\%, which strongly indicates that our optimization policy also works well in the widely-used Autoware software system. The CO policy not only saves significant memory footprint on memory-limited Parker platform but also can potentially improve the timing performance of the GPU-dependent nodes in a real-world case. Similarly, the GPU utilization of CO outperforms MO as shown in Table~\ref{tbl:evalutilization}.

\noindent\textbf{Summary.} Overall, we can observe that the GPU time improvement thanks to CO compared to MO can achieve as high as 34.2\% for micro-benchmarks, 8.06\% for drone case-study, and 20.06\% for Autoware case-study. Since the optimization intuition behind CO is solid and its incurred implementation overhead is trivial, we expect that other real-world workloads may benefit more under CO. Note that MO may incur rather pessimistic latency performance (which may fail the system) under certain scenarios as MO does not consider optimizing latency at all.

%\vspace{-1mm}
\section{Related work}
%\vspace{-1mm}

\noindent\textbf{Discrete GPU Memory Management.} Due to the significant role of data movement in GPU computing, a plenty of work has been done on managing CPU-GPU communication. ~\cite {boyer2013load, conti2012gpu, delorme2013parallel, zhou2015gpes, liu2012power, zhou2018s, hetherington2012characterizing,lee2013performance, spafford2012tradeoffs, 10.1145/2967938.2967944, 8326996} evaluated data transfer overhead and made a comparison between discrete and integrated GPU system.
Apart from managing data movement, much effort has also been dedicated to unifying CPU-GPU memory space to decrease programming burden and increase system performance~\cite{pichai2014architectural, vesely2016observations, garcia2017memory, nielsen2000unified}. 
Nvidia began to support Unified Virtual Memory (UMA) from CUDA 4.0 and some new-types GPU structure~\cite{cook2012cuda}. Analysis of data access pattern and data transfer performance in CUDA  under the UMA mechanism have also been proposed~\cite{landaverde2014investigation, negrut2014unified}.

\noindent\textbf{Integrated GPU Memory Management.} \cite{cudaTegra} summarizes the main charateristics of each memory management method on Tegra platform, but it doesn't quantitatively detail the overhead of each method. Dashti et al.~\cite{dashti2017analyzing} compared the performance of applications which adopt different programming frameworks under the unified memory mechanism on integrated GPU system, and also proposed a new management policy $sharedalloc()$, which allows programmers to manually flush cache. However, this new policy also increases the burden of programming. 
Otterness et al.~\cite{otterness2017evaluation} did a simple comparison among the three GPU MM policies in CUDA, and indicated that only in some scenarios can unified memory benefit the application. 
Li et al.~\cite{li2015evaluation} measured the performance loss of unified memory in CUDA on both integrated and discrete GPU systems and explored the underlying reasons. 
~\cite{mukherjee2016comprehensive} characterized the benefits when unified memory is introduced in OpenCL 2.0 and heterogeneous system architecture on AMD platforms. 
Hestness, Keckler and Wood ~\cite{hestness2015gpu} analyzed the potential opportunities to optimize computing and cache efficiency  on integrated GPU architecture, and proposed a solution for supporting hardware coherence between CPUs and GPUs~\cite{power2013heterogeneous}.

\section{Acknowledgment}
We appreciate our shephard and anonymous reviewers for their valuable suggestions. This work is supported in part by NSF grants CNS 1527727, CNS 1750263 (CAREER), CCF 1822985, and CCF 1943490 (CAREER).

%\vspace{-1mm}
\section{Conclusion}
%\vspace{-1mm}

In this paper, we established a memory management framework for the integrated CPU/GPU architecture, with the goal of assigning the best GPU MM policy for applications that yields optimized response time performance while minimizing the GPU memory footprint.
Our extensive evaluation using both benchmarks and real-world case studies on several platforms proves the accuracy of the proposed performance model and efficacy of the overall approach.
\bibliographystyle{IEEEtran}
\bibliography{refs}
%%%%%%%%%%%%%%%%%%%%%%%%%%%%%%%%%%%%

\end{document}